\documentclass[article,twocolumn]{aastex62}
\usepackage{lineno}
\usepackage{soul}

\newcommand{\LVC}{LIGO/Virgo}

\newcommand{\GBM}{GBM}
\newcommand{\Fermi}{\emph{Fermi}}
\newcommand{\FermiGBM}{\emph{Fermi}-GBM}
\newcommand{\PyCBC}{{\tt PyCBC}}
\newcommand{\GstLAL}{{\tt GstLAL}}
\newcommand{\RAVEN}{\textsc{Raven}}
\newcommand{\BAYESTAR}{\textsc{Bayestar}}

\newcommand{\logLR}{\Lambda}

\newcommand{\Msun}{M$_\odot$}

\begin{document}

\title{A Joint \FermiGBM\ and \LVC\ Analysis of Compact Binary Mergers From the First and Second Gravitational-wave Observing Runs}

%
%
%
%
\author{R.~Hamburg}
\affiliation{Department of Space Science, University of Alabama in Huntsville, Huntsville, AL 35899, USA}
\affiliation{Center for Space Plasma and Aeronomic Research, University of Alabama in Huntsville, Huntsville, AL 35899, USA}
\author{C.~Fletcher}
\affiliation{Science and Technology Institute, Universities Space Research Association, Huntsville, AL 35805, USA}
\author{E.~Burns}\thanks{NASA Postdoctoral Fellow}
\affiliation{NASA Goddard Space Flight Center, Greenbelt, MD 20771, USA}
\author[0000-0002-0587-7042]{A.~Goldstein}
\affiliation{Science and Technology Institute, Universities Space Research Association, Huntsville, AL 35805, USA}
%
%
%
\author{E.~Bissaldi}
\affiliation{Dipartimento Interateneo di Fisica, Politecnico di Bari, Via G. Amendola 126, 70126, Bari, Italy}
\affiliation{Istituto Nazionale di Fisica Nucleare (INFN) Sezione di Bari, Via E. Orabona 4, 70125, Bari, Italy}
\author{M.~S.~Briggs}
\affiliation{Department of Space Science, University of Alabama in Huntsville, Huntsville, AL 35899, USA}
\affiliation{Center for Space Plasma and Aeronomic Research, University of Alabama in Huntsville, Huntsville, AL 35899, USA}
\author{W.~H.~Cleveland}
\affiliation{Science and Technology Institute, Universities Space Research Association, Huntsville, AL 35805, USA}
\author{M.~M.~Giles}
\affiliation{Jacobs Space Exploration Group, Huntsville, AL 35806, USA}
\author{C.~M.~Hui}
\affiliation{NASA Marshall Space Flight Center, Huntsville, AL 35812, USA}
\author{D.~Kocevski}
\affiliation{NASA Marshall Space Flight Center, Huntsville, AL 35812, USA}
\author{S.~Lesage}
\affiliation{Department of Space Science, University of Alabama in Huntsville, Huntsville, AL 35899, USA}
\affiliation{Center for Space Plasma and Aeronomic Research, University of Alabama in Huntsville, Huntsville, AL 35899, USA}
\author{B.~Mailyan}
\affiliation{Center for Space Plasma and Aeronomic Research, University of Alabama in Huntsville, Huntsville, AL 35899, USA}
\author[0000-0002-0380-0041]{C.~Malacaria}\thanks{NASA Postdoctoral Fellow}
\affiliation{NASA Marshall Space Flight Center, Huntsville, AL 35812, USA}
\affiliation{Science and Technology Institute, Universities Space Research Association, Huntsville, AL 35805, USA}
\author[0000-0002-6269-0452]{S.~Poolakkil}
\affiliation{Department of Space Science, University of Alabama in Huntsville, Huntsville, AL 35899, USA}
\affiliation{Center for Space Plasma and Aeronomic Research, University of Alabama in Huntsville, Huntsville, AL 35899, USA}
\author{R.~Preece}
\affiliation{Department of Space Science, University of Alabama in Huntsville, Huntsville, AL 35899, USA}
\author{O.~J.~Roberts}
\affiliation{Science and Technology Institute, Universities Space Research Association, Huntsville, AL 35805, USA}
\author[0000-0002-2149-9846]{P.~Veres}
\affiliation{Center for Space Plasma and Aeronomic Research, University of Alabama in Huntsville, Huntsville, AL 35899, USA}
\author[0000-0002-0221-5916]{A.~von Kienlin}
\affil{Max-Planck-Institut f\"{u}r extraterrestrische Physik, Giessenbachstrasse 1, D-85748 Garching, Germany}
\author[0000-0002-8585-0084]{C.~A.~Wilson-Hodge}
\affiliation{NASA Marshall Space Flight Center, Huntsville, AL 35812, USA}
\author{J.~Wood}\thanks{NASA Postdoctoral Fellow}
\affiliation{NASA Marshall Space Flight Center, Huntsville, AL 35805, USA}
\collaboration{\Fermi\ Gamma-Ray Burst Monitor}



\author{R.~Abbott}
\affiliation{LIGO, California Institute of Technology, Pasadena, CA 91125, USA}
\author{T.~D.~Abbott}
\affiliation{Louisiana State University, Baton Rouge, LA 70803, USA}
\author{S.~Abraham}
\affiliation{Inter-University Centre for Astronomy and Astrophysics, Pune 411007, India}
\author{F.~Acernese}
\affiliation{Dipartimento di Farmacia, Universit\`a di Salerno, I-84084 Fisciano, Salerno, Italy}
\affiliation{INFN, Sezione di Napoli, Complesso Universitario di Monte S.Angelo, I-80126 Napoli, Italy}
\author{K.~Ackley}
\affiliation{OzGrav, School of Physics \& Astronomy, Monash University, Clayton 3800, Victoria, Australia}
\author{C.~Adams}
\affiliation{LIGO Livingston Observatory, Livingston, LA 70754, USA}
\author{R.~X.~Adhikari}
\affiliation{LIGO, California Institute of Technology, Pasadena, CA 91125, USA}
\author{V.~B.~Adya}
\affiliation{OzGrav, Australian National University, Canberra, Australian Capital Territory 0200, Australia}
\author{C.~Affeldt}
\affiliation{Max Planck Institute for Gravitational Physics (Albert Einstein Institute), D-30167 Hannover, Germany}
\affiliation{Leibniz Universit\"at Hannover, D-30167 Hannover, Germany}
\author{M.~Agathos}
\affiliation{Theoretisch-Physikalisches Institut, Friedrich-Schiller-Universit\"at Jena, D-07743 Jena, Germany}
\affiliation{University of Cambridge, Cambridge CB2 1TN, UK}
% ...

\collaboration{The LIGO Scientific Collaboration and the Virgo Collaboration}


\date{\today}

\begin{abstract}
%
%
We present results from offline searches of \Fermi\ Gamma-ray Burst Monitor (GBM) data for gamma-ray transients coincident with the compact binary coalescences observed by the gravitational-wave (GW) detectors Advanced LIGO and Advanced Virgo during their first and second observing runs.
In particular, we perform follow-up for both confirmed events and low significance candidates reported in the \LVC\ catalog GWTC-1.
We search for temporal coincidences between these GW signals and GBM triggered gamma-ray bursts (GRBs).
We also use the GBM Untargeted and Targeted subthreshold searches to find coincident gamma-rays below the on-board triggering threshold.
This work implements a refined statistical approach by incorporating
GW astrophysical source probabilities and GBM visibilities of LIGO/Virgo sky localizations to search for cumulative signatures of coincident subthreshold gamma-rays.
All search methods recover the short gamma-ray burst GRB 170817A occurring $\sim$1.7 s after the binary neutron star merger GW170817.
We also present results from a new search seeking GBM counterparts to LIGO single-interferometer triggers. 
This search finds a candidate joint event, but given the nature of the GBM signal and localization, as well as the high joint false alarm rate of $1.1 \times 10^{-6}$ Hz, we do not consider it an astrophysical association.
We find no additional coincidences.\\
\end{abstract}
%
\section{Introduction}\label{sec:intro}
%
%
Simultaneous observations of the same source in gravitational waves (GWs) and gamma-rays probe some of the most cataclysmic events in the Universe and create rich opportunities to study fundamental physics, cosmology, and high energy astrophysics.
This was demonstrated by the joint observations~\citep{Abbott_2017} of the binary neutron star (BNS) coalescence GW170817~\citep{PhysRevLett.119.161101,PhysRevX.9.011001}
and the short gamma-ray burst GRB 170817A~\citep{Goldstein_2017b,Savchenko_2017}.
These observations led to constraints on the speed of gravity~\citep{abbott2017gravitational}, an independent measure of the Hubble constant~\citep{2017Natur.551...85A, 2019NatAs.tmp..385H, collaboration2019gravitationalwave}, evidence for heavy element production via r-process nucleosynthesis in a kilonova (e.g., \citealt{Chornock_2017,Cowperthwaite_2017,2017Natur.551...80K,Tanvir_2017,Watson_2019}), and more.
Motivated by the wealth of science gained from multi-messenger observations such as these, we seek to increase the number of joint GW/gamma-ray detections by performing coordinated analysis of candidates from Advanced LIGO~\citep{TheLIGOScientific:2014jea}, Advanced Virgo~\citep{TheVirgo:2014hva}, and the \Fermi\ Gamma-ray Burst Monitor (GBM;~\citealt{Meegan_2009}).

The first LIGO/Virgo science observing run (O1) ran from September 2015 to January 2016, during which GBM performed online analyses of GW candidates from compact binary coalescence (CBC) searches.
For GBM offline analysis \citep{Burns_2019}, trigger selection was conservative, treating all CBC candidates with a false alarm rate (FAR) of less than $10^{-5}$~Hz (about 1/day) as equally plausible for follow-up.
The CBC candidates were used to search for coincidences with GBM-triggered GRBs and subthreshold short GRBs from the offline Untargeted Search (Briggs et al., in prep.).
CBC event times were also used to seed more sensitive follow-up with the Targeted Search \citep{Blackburn2015} of \GBM\ data.
No unambiguous coincidences were found between the \GBM\ and \LVC\ candidates.
The most significant event found in the \GBM\ follow-up search was associated with the first observed binary black hole (BBH) coalescence, GW150914 \citep{PhysRevLett.116.061102}.
However the \GBM\ candidate, GW150914-GBM, could not be unambiguously claimed as an electromagnetic counterpart due to its extremely weak signal and poor localization \citep{Connaughton_2016, Greiner_2016, Connaughton_2018}.

For the second observing run (O2), running from November 2016 to August 2017, the \GBM\ Targeted Search was improved \citep{ goldstein2016updates} and run autonomously, in low latency, again following up CBC triggers with $\rm FAR< 10^{-5}$~Hz.
The most interesting multimessenger event from O2 was the association between GW170817 and GRB 170817A.
The Targeted Search proved redundant in this case, as the GRB produced a trigger onboard \Fermi.\footnote{\url{https://gcn.gsfc.nasa.gov/other/524666471.fermi}}
However, had the source been $\sim$10~Mpc farther from Earth, it would not have triggered the detectors onboard GBM and would have only been detectable with subthreshold searches \citep{Abbott_2017, Goldstein_2017b}, while still being well within the \LVC\ detection horizon \citep{PhysRevLett.119.161101}.

In this work, we perform an offline follow-up of all CBC triggers published in the first LIGO/Virgo gravitational-wave transient catalog (GWTC-1; \citealt{2018arXiv181112907T}).
Our search methods are akin to LIGO/Virgo searches for GWs coincident with GRBs \citep{Abbott_2017_150906B,Abbott_2019}.
In addition to seeking coincidences to individual GW events, we search on a statistical basis, looking for any cumulative effects that subthreshold gamma-ray counterparts might have on the resulting follow-up distribution.
We improve upon the GBM analysis of O1 triggers in \citet{Burns_2019}, in that the joint association calculation no longer treats all CBC candidates equally.
Instead, the analysis accounts for the astrophysical nature of the CBC candidates as well as their potential visibility with respect to GBM. This is done by incorporating the probability that each CBC candidate originated from an astrophysical rather than terrestrial source and also considering the fraction of LIGO/Virgo localization probability that was observable to \GBM\ at GW trigger time.
Finally, we augment GBM follow-up of GW events by also reporting results from a new search method \citep{stachie2020search} that seeks gamma-rays coincident with LIGO single-interferometer triggers.

This paper is organized as follows.
In Section \ref{sec:method}, we describe the sample of gravitational-wave candidates and the GBM searches used to follow-up this sample.
Section \ref{sec:results} summarizes the results of these searches, including the search for coincidences with single-interferometer triggers, and discusses the probability of association between the GW and gamma-ray candidate events. In Section \ref{sec:conclusion}, we conclude and discuss future prospects for GBM follow-up of GWs.
%
\section{Method}\label{sec:method}
%
\subsection{Gravitational-wave Trigger Selection}
%
The Advanced LIGO \citep{TheLIGOScientific:2014jea} and Virgo \citep{TheVirgo:2014hva} observatories are kilometer-scale Michelson laser interferometers designed to detect GWs.
Multiple search pipelines are used to detect CBC events in strain data, with each pipeline making different assumptions about the signals and the detector noise and using different technical solutions to maximize detection efficiency.
We focus on events generated by two pipelines: \PyCBC\ \citep{Usman_2016} and \GstLAL\ \citep{PhysRevD.95.042001}. 
Both rely on accurate physical models of the gravitational waveform radiated by a CBC event and use the models to perform matched filtering on strain data.
The process of matched filtering produces a signal-to-noise ratio (S/N) over a large number of templates covering the CBC parameter space.
The extent of the parameter space chosen for O2 and the method used to construct the template bank are described for \PyCBC\ and \GstLAL\ in \cite{DalCanton:2017ala} and \cite{2018arXiv181205121M}, respectively.
Once the S/N has been calculated over all templates, S/N-peaks above a certain threshold are recorded as single-detector CBC triggers. 
Non-Gaussian and non-stationary detector noise frequently produces non-astrophysical triggers with large S/N, hence the pipelines employ a variety of techniques to veto or down-rank such triggers. 
The surviving triggers are used in a coincidence analysis, and each pair of triggers occurring within the maximum GW travel time between detectors produces a coincident trigger. 
The coincident trigger is assigned a ranking statistic that takes into account (i) S/N in the GW detectors, (ii) signal-based vetoes indicating the compatibility of the waveform with a CBC signal, and (iii) the probability of the observed combination of S/N, time delay, and phase difference at the different detectors to be produced by an astrophysical signal (e.g.~\citealt{Nitz:2017}).
The final step is mapping the coincident rank to a statistical significance, which in the case of CBC pipelines is reported via two different quantities: the FAR of the search at the time of the trigger and the probability that the trigger has an astrophysical origin ($p_{\rm astro}$; \citealt{2019arXiv190306881K}).
$p_{\rm astro}$ is estimated using our current understanding of the population of real signals weighed against the distribution of background (false signals) due to GW detector noise fluctuations.

We perform GBM follow-up of all 25 CBC triggers reported in the LIGO/Virgo catalog GWTC-1 \citep{2018arXiv181112907T}. 
This catalog utilized state-of-the-art configurations of \PyCBC\ and \GstLAL, as well as the best data-quality selection of the LIGO and Virgo strain data available, for a full reanalysis of O1 and O2.
Listed in Table \ref{tab:gw_triggers}, the catalog triggers were required to pass an initial threshold of FAR $\lesssim 3.86 \times 10^{-7}$~Hz (about 1/30 days) in at least one pipeline.
Triggers passing this FAR threshold and additionally having $p_{\rm astro}$ greater than 50\% are denoted with ``GW" in the event name.
In the follow-up analyses, the GBM searches are guided by the CBC trigger times.
To assess GBM coverage of the LIGO/Virgo triggers, the public HEALPix \citep{2005ApJ...622..759G} sky localization maps accompanying GWTC-1 are taken for the high $p_{\rm astro}$ detections \citep{GWTC1_skymaps}.
We generate \BAYESTAR\ skymaps \citep{PhysRevD.93.024013} for all remaining triggers which had corresponding GBM data.
\BAYESTAR\ skymaps rely on the mass and spin parameters reported by the searches and do not marginalize over them, as is done instead for high $p_{\rm astro}$ detections via full parameter estimation \citep{Veitch_2015, Abbott_2016_other}.
Nevertheless, they allow approximations of GBM observing coverages at much lower computational costs.
Finally, for each CBC trigger, the maximum $p_{\rm astro}$ is used between the \GstLAL\ and \PyCBC\ pipelines \citep[][Table IV]{2018arXiv181112907T}.

%
%
\begin{table}[!t]
\begin{center}
\begin{tabular}{c c c c c}
\hline \hline
LIGO/Virgo &   &   &   & GBM \\
GW Event & UTC Date & UTC Time & $p_{\rm astro}$ & Coverage \\
\hline
GW150914 & 2015-09-14 & 09:50:45.4 & 1 & 66.7\% \\
151008 & 2015-10-08 & 14:09:17.5 & 0.27 & 100\% \\
151012.2 & 2015-10-12 & 06:30:45.2 & 0.023 & 58.4\% \\
GW151012 & 2015-10-12 & 09:54:43.4 & 1 & 66.1\% \\
151116 & 2015-11-16 & 22:41:48.7 & $\ll$0.5 & 72.6\% \\
GW151226 & 2015-12-26 & 03:38:53.6 & 1 & 78.8\% \\
161202 & 2016-12-02 & 03:53:44.9  & 0.034 & - \\
161217 & 2016-12-17 & 07:16:24.4 & 0.018 & - \\
GW170104 & 2017-01-04 & 10:11:58.6 & 1 & 90.3\% \\
170208 & 2017-02-08 & 10:39:25.8 & 0.02 & 97.8\% \\
170219 & 2017-02-19 & 14:04:09.0 & 0.02 & 5.1\% \\
170405 & 2017-04-05 & 11:04:52.7 & 0.004 & - \\
170412 & 2017-04-12 & 15:56:39.0 & 0.06 & 67.2\% \\
170423 & 2017-04-23 & 12:10:45.0 & 0.086 & 45.2\% \\
GW170608 & 2017-06-08 & 02:01:16.5 & 1 & 73.0\% \\
170616 & 2017-06-16 & 19:47:20.8 & $\ll$0.5 & 66.2\% \\
170630 & 2017-06-30 & 16:17:07.8 & 0.02 & 8.2\% \\
170705 & 2017-07-05 & 08:45:16.3 & 0.012 & 26.3\% \\
170720 & 2017-07-20 & 22:44:31.8 & 0.0097 & 48.2\% \\
GW170729 & 2017-07-29 & 18:56:29.3 & 0.98 & 88.9\% \\
GW170809 & 2017-08-09 & 08:28:21.8 & 1 & 73.9\% \\
GW170814 & 2017-08-14 & 10:30:43.5 & 1 & 73.6\% \\
GW170817 & 2017-08-17 & 12:41:04.4 & 1 & 100\% \\
GW170818 & 2017-08-18 & 02:25:09.1 & 1 & 100\% \\
GW170823 & 2017-08-23 & 13:13:58.5 & 1 & - \\
\hline
\end{tabular}
\caption{Gravitational-wave triggers from \cite{2018arXiv181112907T}. The $p_{\rm astro}$ values shown here are the maximum values reported between the \GstLAL\ and \PyCBC\ pipelines.
The percentage of the LIGO/Virgo localization probability that was visible to GBM at trigger time is also given.
Triggers with unspecified coverage are due to GBM passage through the South Atlantic Anomaly when all detectors are turned off. 
}
\label{tab:gw_triggers}
\end{center}
\end{table}

%
\subsection{\FermiGBM\ Searches}

GBM is a survey instrument aboard the \Fermi\ Gamma-ray Space Telescope and is comprised of 14 scintillation detectors that span an energy range of 8~keV to 40~MeV \citep{Meegan_2009}.
Twelve of the detectors are made of thallium-doped sodium iodide (NaI) crystals and are oriented in such a manner as to cover the entire sky un-occulted by the Earth ($\sim$70\%).
The two other detectors are bismuth germanate (BGO) crystals positioned on opposite sides of the spacecraft.
Triggering algorithms running on the satellite search data on multiple timescales and energy ranges for coherent, statistically-significant (usually 4$\sigma$) excesses in at least 2 NaI detectors (\citealt{GBMBurstCatalog_6Years}; von Kienlin et al., in prep.).
Localization is performed by combining the detector responses with a set of three template photon spectra representing spectrally hard, normal, and soft GRBs to generate expected photon counts from points evenly spaced across a 1$^\circ$ grid of the sky \citep{Connaughton_2015}.
The expected count rates are compared to the observed rates, and a $\chi^2$ minimization process identifies the most likely direction, with localization accuracy on the order of degrees.
GBM continuously takes data except during passage through the South Atlantic Anomaly (SAA) when the detectors are turned off due to high particle flux, yielding an uptime of approximately 85\%.

GBM has developed increased sensitivity to weak, short GRBs by means of two offline searches: the Untargeted Search\footnote{\url{https://gcn.gsfc.nasa.gov/fermi_gbm_subthresh_archive.html}} (Briggs et al., in prep.) and the Targeted Search \citep{Blackburn2015, goldstein2016updates}.
These searches seek transient signals that do not exceed the high threshold set by the on-board triggering algorithms, and in this work, they are employed to find subthreshold gamma-rays coincident with the GW triggers in our search sample. 
Additional details on these searches follow.

\subsubsection{Untargeted Search}
The Untargeted Search is a blind search of continuous time-tagged event (CTTE) data, running automatically upon receipt of data from the \Fermi~spacecraft and using no information from GW searches.
The search improves upon the onboard triggering algorithms by utilizing additional energy ranges and timescales, as well as a more sophisticated background-fitting model.
Candidate events are required to have excess counts greater than 2.5$\sigma$ relative to background in one detector and at least 1.25$\sigma$ in a second detector.
Significant candidates are autonomously distributed via the Gamma-ray Coordinates Network along with HEALPix skymaps to facilitate joint detections with other instruments (see e.g., \citealt{grb170921C}).
Further details on the Untargeted Search and an analysis of its candidates will be published in a forthcoming article.

\subsubsection{Targeted Search}
The Targeted Search was designed for multi-messenger follow-up, requiring an input time and/or HEALPix skymap to seed a sensitive search of CTTE data.
When seeking counterparts to GWs, the Targeted Search analyzes a 60 s window centered on the input GW time and searches timescales increasing by powers of 2 from 64~ms to 8.192~s, while phasing time bins by a factor of 4.
Data from all 14 detectors are processed coherently to achieve a greater sensitivity to weak signals than when analyzing one detector at a time, as performed by the on-board flight software and the Untargeted Search.
Three model spectra, described in \cite{goldstein2016updates}, are folded through the detector responses to produce templates of expected counts which are then compared to the observed distribution of counts in each energy channel of each detector. The comparison is performed via a log-likelihood ratio ($\Lambda$), testing the alternative hypothesis of the presence of a signal with a similar spectrum versus the null hypothesis of only background noise.  Treating $\Lambda$ as our detection statistic, the model spectrum resulting in the highest $\Lambda$ is selected as the preferred spectrum, and this procedure is repeated for each bin of data in the search (see \citealt{Blackburn2015} for the detailed calculation of $\logLR$).
Bins contaminated by phosphorescent noise events are removed, and overlapping bins are merged to produce only the most significant bin.
After this filtering, all remaining bins are retained as candidate events for our analysis.
The different spectral templates tend to identify different types of sources in the GBM background, and such types may have very different rates of occurrence.
To preserve sensitivity to these different sources, the bins are separated by best-fit spectral template, and event significance (i.e., FAR) is measured against background from the same template.

The Targeted Search was made more sensitive in preparation for O2 by improving the background estimation, revising the spectral template for hard GRBs, and implementing additional automated filters \citep{goldstein2016updates}.
In particular, a $\logLR$ pre-filter was applied.
The $\logLR$ calculation demands an initial estimation of the signal amplitude (effectively, the photon fluence in the time bin over 50-300 keV) that maximizes the likelihood of the hypothesis that a signal exists.
The pre-filter excludes time bins with initial guesses of $\logLR<5$ from the full numerical optimization, increasing the speed of this computationally expensive task by up to a factor of 5.
Bins with $\logLR < 5$ have been verified to lie well within the GBM background, thus excluding them does not affect the sensitivity of the search.
This updated version of the Targeted Search was used to analyze both the O1 and O2 triggers in our sample.
Further improvements have been made for online analysis of CBC triggers during Advanced LIGO and Advanced Virgo's third observing run \citep{goldstein2019updates}, but were not used in this work.

%
\section{Results}\label{sec:results}
%
%
\begin{figure}[!t]
    \centering
    \includegraphics[width=0.5\textwidth]{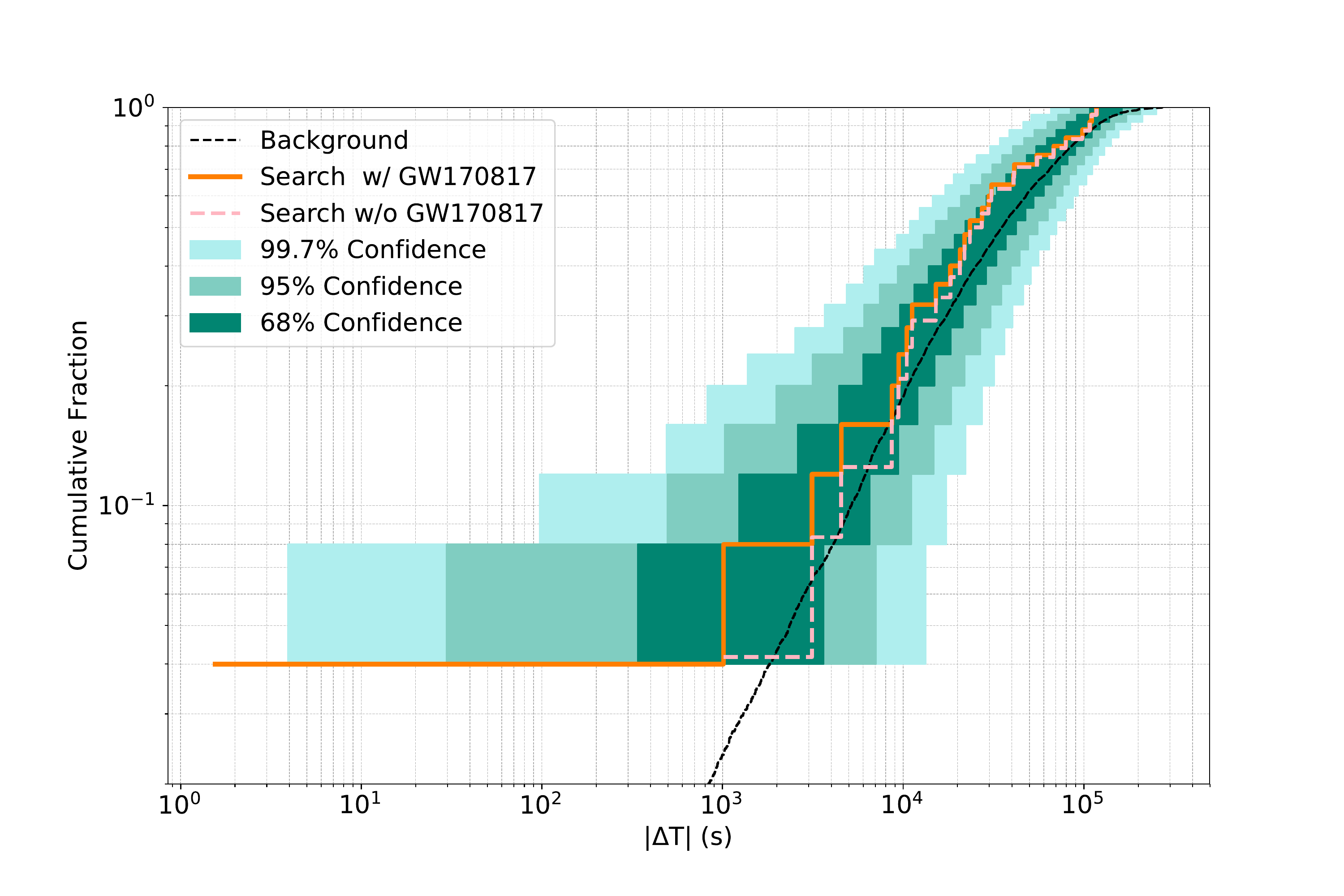}
    \caption{Cumulative distribution for the minimal time offsets between the 25 CBC triggers and GRBs found by either the GBM onboard triggering algorithms or the Untargeted Search.
    The background offset distribution is shown in black. The search sample including GW170817 is depicted by the solid gold line, and the search excluding GW170817 is shown in dashed brown.}
    \label{fig:untarget_trigger_cdf}
\end{figure}
Here we present the results of our searches for gamma-ray counterparts to the GW triggers in our sample.
To quantify event significance, each resulting search distribution is compared to that of background.
The background used in the following sections is composed of randomly selected times during which both LIGO detectors were in observing mode during O1 and O2.
The ratio of random background between O1 and O2 is also roughly proportional to the LIGO/Virgo livetimes during O1 and O2.
The same Targeted Search input parameters used for the search sample were used for the background, resulting in $\sim$10 (20)~ks of background during O1 (O2), yielding a minimum FAR of $\sim 1 \times 10^{-5}$ ($\sim 5 \times 10^{-6}$)~Hz for Targeted Search analysis.
Finally, the background times were chosen independently with respect to GBM and therefore include GBM trigger times.
\subsection{GBM Trigger and Untargeted Search Results}\label{ssec:results_trigger_untargeted}

As done in \citet{Burns_2019}, we first examine the time offsets between the search sample of CBC triggers and both GRBs detected by the GBM on-board flight software and subthreshold short GRB candidates from the Untargeted Search.
This method is similar to the \RAVEN\ analysis used by LIGO/Virgo \citep{UrbanThesis}.
The Untargeted Search sample consists of all 187 candidates published during O1 and O2 via GCN, as described in the previous section. Combining these with the triggered GRBs, we obtained a total of 474 GRBs. The temporal offsets between the 25 GW events and the GBM GRBs were then determined, and the smallest offset for each GW candidate was taken. The search sample offsets are compared to those arising from random coincidences by finding the shortest temporal offsets between the background times and the GW trigger times.
Both positive and negative offsets were allowed for search sample and background, but a maximum offset was not enforced.
GW triggers occurring during \textit{Fermi} passage through SAA were included, limiting the minimum time offsets for some GBM events; however the same treatment for the search was used for background.

The cumulative distribution for this search is presented in Figure~\ref{fig:untarget_trigger_cdf}.
The search sample including GW170817 is shown with the solid gold line, while the distribution without GW170817 is displayed by the dashed brown line.
Confidence regions were obtained empirically by Monte Carlo sampling of the background offset distribution with sample size equal to that of the search sample and finding the desired percentiles.
The most significant deviation of the search distribution from that of random background is caused by GRB 170817A, found $\sim$1.7~s after GW170817.
Omitting GW170817, the shortest time interval between a CBC trigger from our sample and a \GBM\ event is approximately 1000~s.
On-axis prompt emission from a short GRB is not expected at such large time delays after a binary neutron star merger \citep{Vedrenne2009, Zhang2019}, though larger delays may be allowed for off-axis emission (e.g., \citealt{Salafia_2018}).
Hence, with this first search we find no evidence for GW/gamma-ray associations apart from GW170817/GRB 170817A.

\begin{figure*}
\centering
\includegraphics[width=0.9\textwidth]{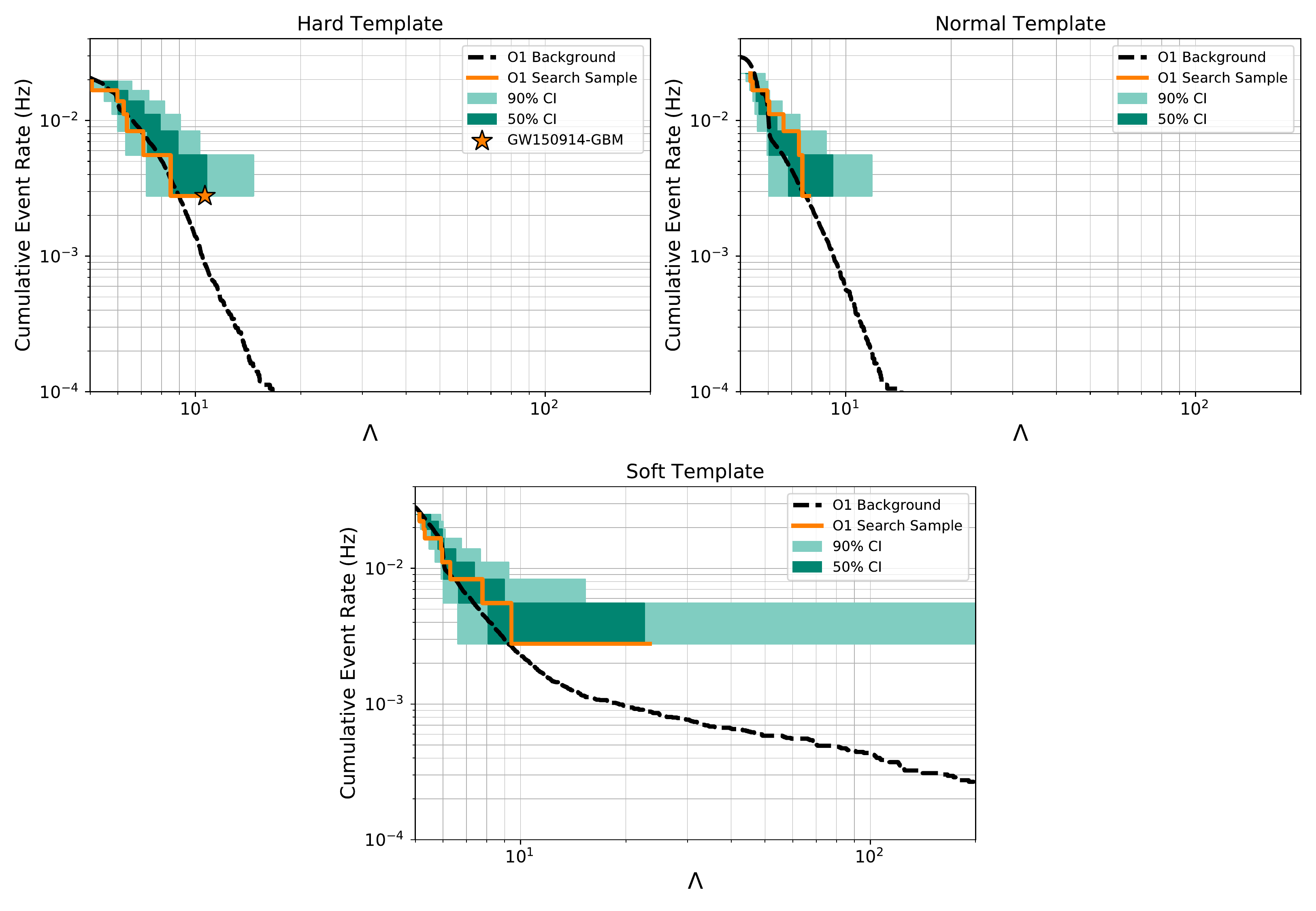}
\caption{O1 cumulative event rate distributions of the GBM background (black dashed lines) and search samples (solid gold line) for the GBM Targeted Search as a function of the log-likelihood ratio. Distributions are separated according to best-fitting spectral template. The transient GW150914-GBM is marked by a gold star in the hard template distribution.}
\label{fig:O1_far_distributions}
\end{figure*}

%
\subsection{Targeted Search Results}\label{ssec:results_targeted}
%
\begin{figure*}
\centering
\includegraphics[width=0.9\textwidth]{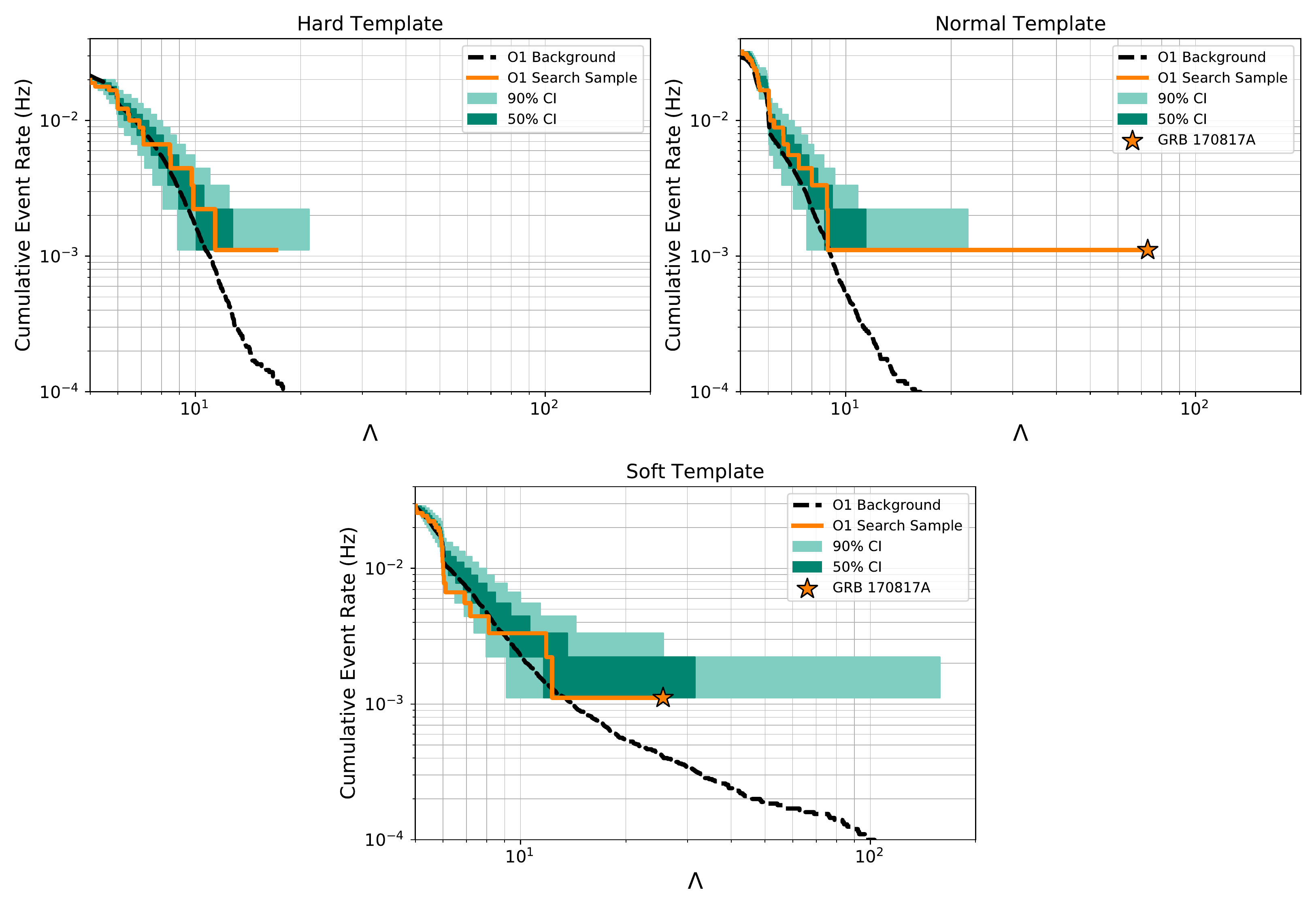}
\caption{O2 cumulative event rate distributions of the  GBM background (black dashed lines) and search samples (solid gold line) for the GBM Targeted Search as a function of the log-likelihood ratio. Distributions are separated according to best-fitting spectral template. Both the main peak and soft thermal tail of GRB 170817A, the short gamma-ray burst counterpart to GW170817, are indicated by gold stars in the normal and soft template distributions, respectively.}
\label{fig:O2_far_distributions}
\end{figure*}
The Targeted Search was used to search for subthreshold gamma-ray signals around 21 events from the CBC search sample.
GBM data were not collected around triggers 161202, 161217, 170405, and GW170823 due to passage through the SAA; therefore these events were excluded from this search.
For those remaining, the GBM coverage of the LIGO/Virgo
localizations (see Table \ref{tab:gw_triggers}) was obtained.
No LIGO/Virgo skymap was fully occulted by the Earth, and GBM observed between $\sim$5\% and 100\% of the localization probability with an average observing fraction of 67.0\%.

The Targeted Search search follow-up distributions for O1 triggers and O2 triggers are shown as functions of $\logLR$ in Figures \ref{fig:O1_far_distributions} and \ref{fig:O2_far_distributions}, respectively.
The background distributions were constructed by running the Targeted Search over the randomly selected times described above with the same parameters used for the search sample. 
As described in the previous section, confidence intervals for the search samples were produced by Monte Carlo sampling the background $\logLR$ distributions with the same sample size as the search sample.
The distributions are separated into three categories according to the best-fitting spectral template, due to the different backgrounds affecting the three templates.
Also, because of the time-variable nature of the background in each template, we obtain event significance by comparing the follow-up of O1 triggers to GBM background taken during O1 and O2 follow-up to O2 background.

For both O1 and O2, the search distributions lie largely within the 90\% confidence region of the median for all spectral templates.
The O1 follow-up (Figure \ref{fig:O1_far_distributions}) does not show any significant outliers in the sample distributions. 
The transient GW150914-GBM is found with a FAR of $8.7 \times 10^{-4}$~Hz in the hard template distribution, where the FAR is the cumulative event rate of the background at the same $\logLR$, and lies just within 50\% confidence.
The most significant event in the O2 follow-up (Figure \ref{fig:O2_far_distributions}) can be seen in the normal template distribution and is GRB 170817A, found with a FAR of $2.0 \times 10^{-5}$~Hz.
The spectrally soft tail of GRB 170817A is also the most significant foreground event in the O2 soft template distribution, with a FAR of $4.1 \times 10^{-4}$~Hz, but is within the 50\% confidence region.
No other significant candidates are found.

\subsection{Targeted Search Joint Analysis}\label{ssec:joint_analysis}
The FARs discussed in the previous section measure the significance of GBM transients with respect to the Targeted Search background only, regardless of the GW observations. 
Here we characterize the significance of coincidences between the GW events and the gamma-ray signals from the Targeted Search.
In our previous works (e.g., \citealt{Connaughton_2016, Burns_2019}), this was done by ranking gamma-ray candidates by the Targeted Search FAR and the relative time offsets between the candidates and the GW triggers.
We build upon these analyses by also considering (i) the probability that the GW signal is astrophysical in origin and (ii) the fraction of the LIGO/Virgo sky localization visible to GBM at the GW event time.
Therefore, we rank gamma-ray candidates found by the Targeted Search with a statistic $R$ defined as 

\begin{equation}
   R = \frac{p_{\rm astro} \times p_{\rm visible}} {|\rm \Delta \textit{t}| \times \text{FAR}_{\rm GBM}},
\end{equation}

\noindent where $\Delta t$ is the time offset between the GW trigger and the gamma-ray event and $p_{\rm visible}$ is the fraction of the LIGO/Virgo localization probability observable to GBM.
A minimum offset of 64~ms was set to match the time binning of the data.
GW triggers 151116 and 170616 were given the lowest $p_{\rm astro}$ of the sample (i.e., 0.004) in light of the upper limits reported in GWTC-1 (see Table \ref{tab:gw_triggers}).
Background events are ranked using the same statistic $R$.
As background events have no corresponding LIGO/Virgo information, skymaps and $p_{\rm astro}$ values from the GW search sample were randomly assigned to each background event, and the fraction of GBM visibility was calculated at the background time using the randomly-selected skymap.

\begin{figure}
   \begin{minipage}{0.5\textwidth}
     \centering
     \includegraphics[width=1\linewidth]{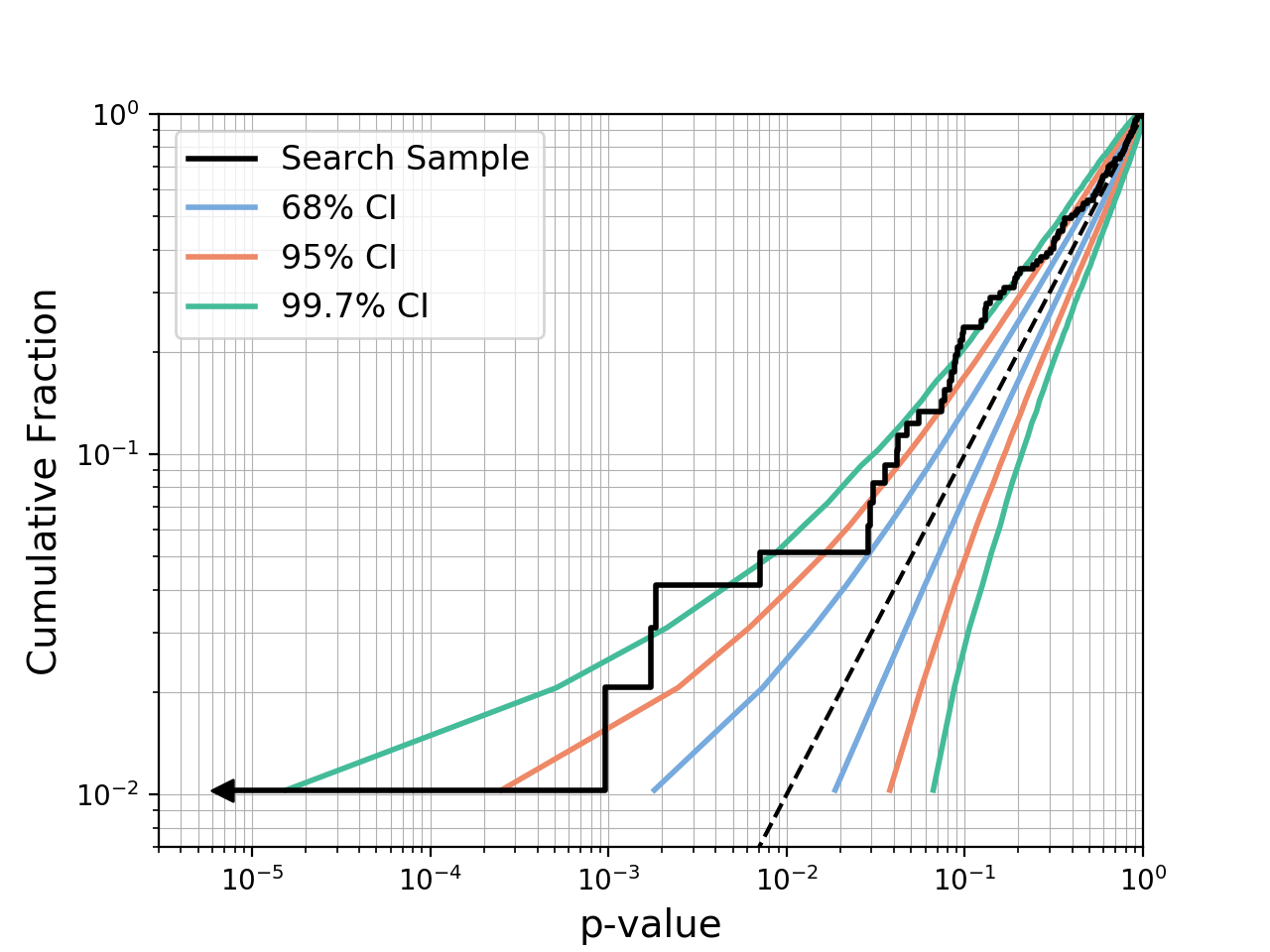}
   \end{minipage}\hfill
   \begin{minipage}{0.5\textwidth}
     \centering
     \includegraphics[width=1\linewidth]{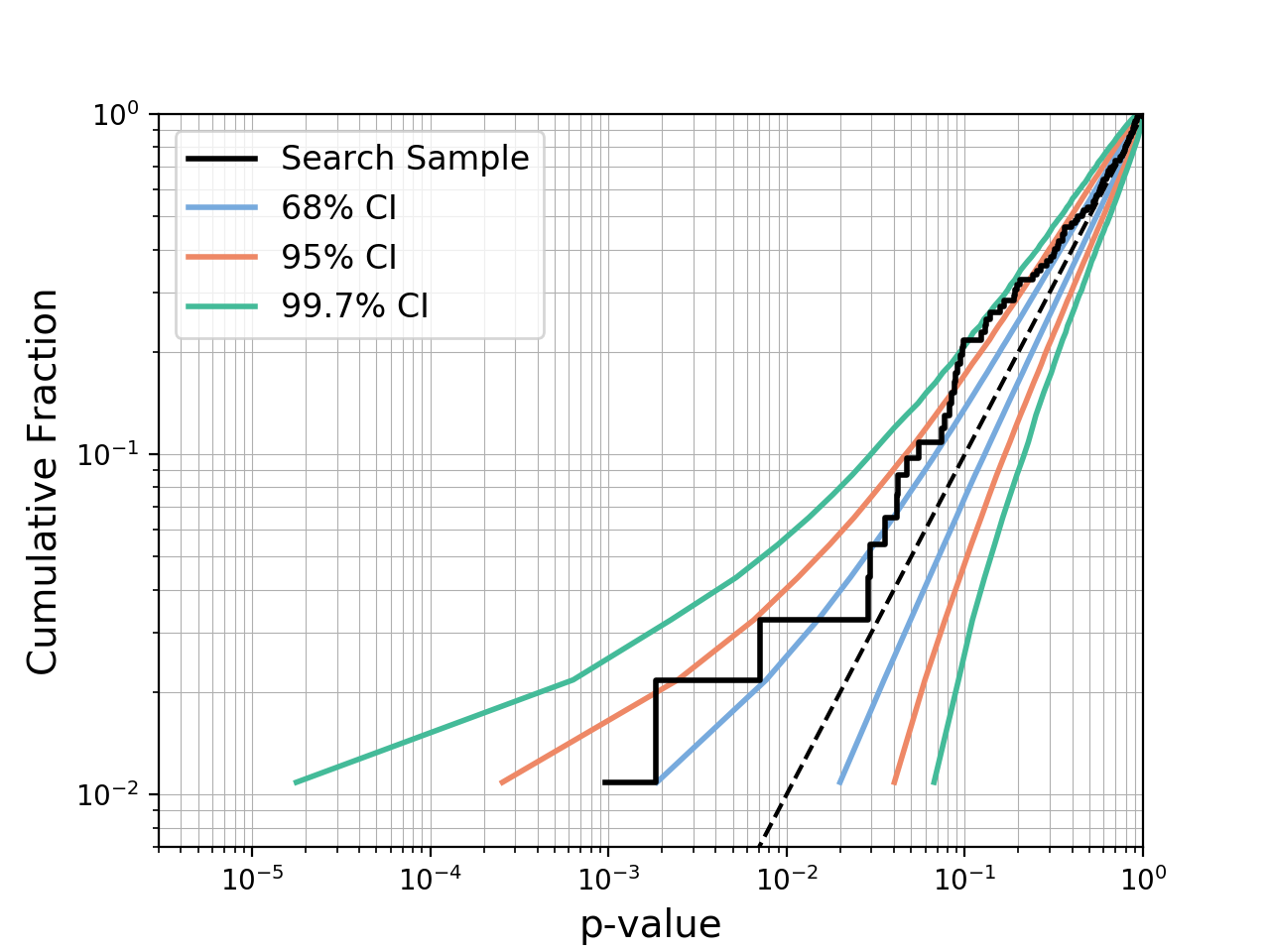}
   \end{minipage}
\caption{Cumulative distribution of the Targeted Search \mbox{p-values}. The dashed black lines represent the expected background distribution. Top: Follow-up search sample including GW170817.  The main emission episode of GRB 170817A is found with higher ranking than any other candidate within the background distribution. Its p-value is therefore marked as an upper limit (black triangle) at greater than 3$\sigma$ deviation from the background p-value distribution. Bottom: Follow-up search sample without GW170817.}
    \label{fig:joint_pvalues}
\end{figure}

The ranking statistic of the search sample is mapped to a p-value, defined as the number of more highly ranked background events divided by the total number of background events, or $p_i$ = $N(R > R_i)/N$, where $N$ is the number of gamma-ray events in the background and $i$ is the index of an event in the search sample.
Again, search sample events from O1 and O2 are compared to background from O1 and O2, respectively.
The cumulative distributions of the combined O1 and O2 p-values are shown in Figure \ref{fig:joint_pvalues}, with and without GW170817 follow-up.
The dashed black lines follow a uniform distribution, representing the null hypothesis that the search sample is consistent with that of background.
The confidence regions for the p-value distribution were generated by random sampling of the background uniform distribution with sample size equal to the search sample size.

For the search including GW170817 follow-up, excesses of greater than $3\sigma$ are observed due to contributions from GRB 170817A. 
The main emission peak of GRB 170817A has a higher ranking than any other event in the background, making its p-value an upper limit.
Removing all Targeted Search candidates associated with GW170817, excesses greater than 2$\sigma$ are still observed.
Contributing to this near the tail of the distribution is GW150914-GBM, which is found with a p-value of $\sim1.8 \times 10^{-3}$.
Of the remaining candidates (located around p-value = $1.0 \times 10^{-1}$), the detector lightcurves, spectral information, and localizations have been manually inspected.
Real signals have consistent signal in detectors viewing approximately the same portion of the sky and are likely be found on multiple timescales by the Targeted Search. Short GRB-like signals typically display most of their emission above 50~keV. 
However, softer events with localizations consistent the Sun or the Galactic plane are likely to be solar flares or galactic sources rather than GRBs.
All inspected events were judged to be either inconsistent with real short GRB-like signals or too weak in GBM data to constrain any properties. 
Therefore we judge this excess likely unrelated to the CBCs in the search sample.
Some of the excess may be due to real but unrelated gamma-ray signals, and
future observations can be used to either exclude or strengthen this feature.
We do not find evidence here to report any associations other than GW170817 and GRB 170817A.

%
\subsection{Targeted Search Follow-up of Single Interferometer Triggers}\label{ssec:results_single_ifo}
%
%
During O1 and O2, a single LIGO interferometer taking science observing-mode data covered $33.4\%$ and $29.5\%$ of the respective livetimes.
CBC events occurring during these times can still be detected \citep{Callister:2017urp, Sachdev:2019vvd}, albeit with a reduced significance due to the lack of coincidence with a second detector.
The lack of a second detector can be somewhat mitigated by searching for a coincident gamma-ray transient~\citep{Nitz_2019} as the physical connection between GWs and GRBs has been established for at least BNS mergers.
This idea is roughly illustrated by the narrative of GW170817, which was initially a single interferometer trigger due to the presence of a glitch in the LIGO Livingston detector~\citep{PhysRevLett.119.161101, 2018PhRvD..98h4016P}, but was nonetheless found to be time-coincident with GRB 170817A.

The method for searching for GBM counterparts to single-interferometer triggers differs from those presented in the previous sections.
We start from \PyCBC\ single-interferometer triggers having a reweighted S/N \citep{Usman_2016} higher than 8, yielding a sample of 1621 (1126 for O2 and 495 for O1) triggers.
The search for gamma-ray counterparts is then performed using the Targeted Search. 
We only consider possible associations between \PyCBC\ candidates and the most significant GBM candidates found within the corresponding $\pm$ 30~s search windows.
Thus, we obtain pairs of GW candidates and gamma-ray candidates and compute a joint statistical significance. 
This statistic is calculated by taking into account (i) the time offset, (ii) the reweighted S/N of the GW trigger, (iii) the Targeted Search $\logLR$, and (iv) the overlap between the GW and gamma-ray sky localizations defined in \cite{Ashton_2018}. 
Further details on the statistical method will be given in \cite{stachie2020search}.
Although we find no highly significant associations, a close inspection of the data around the 80 candidates with the highest significance (i.e. lowest FAR) was performed. 
For these candidates, LIGO detector characterization was performed using standard tools like Omicron scans, Omega scans, and Used Percentage Veto~\citep{abbott2018effects, Abbott_2016, Isogai_2010}.
Sixty-four candidates in temporal proximity with known types of instrumental transients, blip glitches \citep{Abbott_2016, cabero2019blip}, non-stationary noise visible in spectrograms, and scattered light were rejected. 
There were 12 other triggers disfavoured because parameter estimation \citep{PhysRevD.81.062003} either showed evidence of a glitch (i.e., the existence of bimodality in posterior probability for different CBC parameters) or returned a low ($<5$) $\text{log}_{10}$ Bayes factor. The Bayes factor compares the hypothesis of the presence of signal in the data to the hypothesis of the presence of Gaussian noise, with a low Bayes factor indicating the data contain little evidence of a signal. Three candidates were also eliminated due to noticeably poor background fits in the low-energy channels of the GBM detectors, which often cause inflated $\logLR$ values.

A single L1 surviving coincident association remained with no obvious reason for rejection. 
However, the derived FAR, based on coincidences between noises in LIGO and noises in GBM \citep{stachie2020search}, is relatively high at $1.1 \times 10^{-6}$~Hz. The implied low significance is mainly due to the soft spectrum of the GBM candidate. 
The GBM candidate has a localization consistent with the galactic plane and is likely produced by Scorpius X-1, as a strong occultation step caused by this Galactic X-ray source was observed close in time to the trigger.
Finally, the parameter estimation of the LIGO signal indicates masses of $>100$~\Msun\ for the two components of the binary. As of yet, there are no confirmed observations of such binary mergers~\citep{PhysRevD.100.064064}, which suggests that these systems, if they exist, are not common.

%
%
\section{Summary and Future Directions}\label{sec:conclusion}
%
We have used LIGO/Virgo and \Fermi-GBM data and multiple algorithms to search for gamma-ray transients associated with high and low significance CBC events reported in the first gravitational-wave transient catalog, GWTC-1. 
The GBM subthreshold searches for gamma-ray candidates employed improved algorithms to conduct more sensitive searches than those used in online follow-up during O1 and O2.
All searches identified the coincidence between the short gamma-ray burst GRB 170817A and the BNS coalescence signal GW170817. 
We found no additional coincident detections between CBC triggers and GBM triggers or Untargeted Search candidates.
The GBM Targeted Search found the main emission peak and the long, soft tail of GRB 170817A with FARs of $2.0 \times 10^{-5}$~Hz and $4.1 \times 10^{-4}$~Hz, respectively, and the p-value of the joint association was found to deviate from the background distribution at greater than 3 sigma.
The gamma-ray transient GW150914-GBM was also found with a FAR of $8.7 \times 10^{-4}$~Hz, but was not a significant candidate on its own, lying just within the 50\% confidence region of the hard spectral template.
Future multi-messenger observations will be necessary to establish any astrophysical connection between gamma-ray emission and BBH mergers (see e.g., \citealt{Veres:2019hsd}).
No other short GRB candidates were found in association with the CBC triggers.

In this work, the joint analysis was improved compared to that performed in \citet{Burns_2019}.
In addition to the temporal offset and the Targeted Search FAR, we also considered the significance of the LIGO/Virgo trigger and the GBM visibility of the LIGO/Virgo sky localization.
However, this analysis can be further refined.
By including all candidates reported in GWTC-1, we implicitly assumed that BBH, BNS, and NSBH (i.e., neutron star-black hole) mergers are equally likely to produce gamma-ray emission, and sought counterparts to these mergers using a wide parameter space of different timescales, energy ranges, and spectral templates.
The broad nature of this search was motivated by the fact that, with only one confirmed coincidence, the observational properties of joint GW/GRB events are still largely unknown.
Improving our search to target short GRB-like signals and filter transients from sources unrelated to CBCs, such as particle and galactic flares, may increase sensitivity to coincident, subthreshold short GRBs.
Improvements in GBM search pipelines \citep{goldstein2019updates} and formal methodology (e.g., \citealt{Ashton_2018}) are being undertaken for joint LIGO/Virgo and GBM analysis of CBC triggers from O3.

Finally, a new search for GBM coincidences with LIGO single-interferometer triggers was also conducted.
The most interesting resulting candidate is unlikely to be an astrophysical association because of its high FAR. 
Additionally, the gamma-ray signal was likely caused by flaring activity from a source near the Galactic plane and parameter estimation of the LIGO signal suggests source masses inconsistent with a neutron-star coalescence.
For future observing runs \citep{collaboration2013prospects}, the single-interferometer search methods will be improved.
The introduction of several types of follow-up methods will be one of the modifications introduced during these subsequent runs.
This will result in an improved FAR distribution, as future observations will assess associations between a specific category of CBC candidates (BNS, NSBH or BBH) and GBM candidates defined by their duration and spectral hardness. \\

The UAH co-authors gratefully acknowledge NASA funding from co-operative agreement NNM11AA01A. The USRA co-authors gratefully acknowledge NASA funding through contract NNM13AA43C. 
Through appointments to the NASA Postdoctoral Program, EB is supported at the Goddard Space Flight Center, and CM and JW are supported at the Marshall Space Flight Center.
CAWH gratefully acknowledges NASA funding through the \Fermi\ GBM project.

The LIGO and Virgo co-authors gratefully acknowledge the support of the United States
National Science Foundation (NSF) for the construction and operation of the
LIGO Laboratory and Advanced LIGO as well as the Science and Technology Facilities Council (STFC) of the
United Kingdom, the Max-Planck-Society (MPS), and the State of
Niedersachsen/Germany for support of the construction of Advanced LIGO 
and construction and operation of the GEO600 detector. 
Additional support for Advanced LIGO was provided by the Australian Research Council.
The authors gratefully acknowledge the Italian Istituto Nazionale di Fisica Nucleare (INFN),  
the French Centre National de la Recherche Scientifique (CNRS) and
the Foundation for Fundamental Research on Matter supported by the Netherlands Organisation for Scientific Research, 
for the construction and operation of the Virgo detector
and the creation and support  of the EGO consortium. 
The authors also gratefully acknowledge research support from these agencies as well as by 
the Council of Scientific and Industrial Research of India, 
the Department of Science and Technology, India,
the Science \& Engineering Research Board (SERB), India,
the Ministry of Human Resource Development, India,
the Spanish  Agencia Estatal de Investigaci\'on,
the Vicepresid\`encia i Conselleria d'Innovaci\'o, Recerca i Turisme and the Conselleria d'Educaci\'o i Universitat del Govern de les Illes Balears,
the Conselleria d'Educaci\'o, Investigaci\'o, Cultura i Esport de la Generalitat Valenciana,
the National Science Centre of Poland,
the Swiss National Science Foundation (SNSF),
the Russian Foundation for Basic Research, 
the Russian Science Foundation,
the European Commission,
the European Regional Development Funds (ERDF),
the Royal Society, 
the Scottish Funding Council, 
the Scottish Universities Physics Alliance, 
the Hungarian Scientific Research Fund (OTKA),
the Lyon Institute of Origins (LIO),
the Paris \^{I}le-de-France Region, 
the National Research, Development and Innovation Office Hungary (NKFIH), 
the National Research Foundation of Korea,
Industry Canada and the Province of Ontario through the Ministry of Economic Development and Innovation, 
the Natural Science and Engineering Research Council Canada,
the Canadian Institute for Advanced Research,
the Brazilian Ministry of Science, Technology, Innovations, and Communications,
the International Center for Theoretical Physics South American Institute for Fundamental Research (ICTP-SAIFR), 
the Research Grants Council of Hong Kong,
the National Natural Science Foundation of China (NSFC),
the Leverhulme Trust, 
the Research Corporation, 
the Ministry of Science and Technology (MOST), Taiwan
and
the Kavli Foundation.
The authors gratefully acknowledge the support of the NSF, STFC, INFN and CNRS for provision of computational resources.

This research also made use of Astropy, a community-developed core Python package for Astronomy \citep{2013A&A...558A..33A}; NumPy \citep{van2011numpy}; SciPy \citep{jones_scipy_2001}; and matplotlib, a Python library for publication quality graphics \citep{Hunter:2007}.
%
%
\bibliographystyle{aasjournal}
\bibliography{main}

\begin{thebibliography}{}
\expandafter\ifx\csname natexlab\endcsname\relax\def\natexlab#1{#1}\fi

\bibitem[{Aasi {et~al.}(2015)}]{TheLIGOScientific:2014jea}
Aasi, J., {et~al.} 2015, Class. Quant. Grav., 32, 074001

\bibitem[{Abbott {et~al.}(2016{\natexlab{a}})Abbott, Abbott, Abbott, Abernathy,
  Acernese, Ackley, Adams, Adams, Addesso, Adhikari, \&
  et~al.}]{Abbott_2016_other}
Abbott, B., Abbott, R., Abbott, T., {et~al.} 2016{\natexlab{a}}, Physical
  Review Letters, 116, doi:10.1103/physrevlett.116.241102

\bibitem[{Abbott {et~al.}(2018)Abbott, Abbott, Abbott, Abernathy, Acernese,
  Ackley, Adams, Adams, Addesso, Adhikari, {et~al.}}]{abbott2018effects}
---. 2018, Classical and Quantum Gravity, 35, 065010

\bibitem[{Abbott {et~al.}(2019{\natexlab{a}})Abbott, Abbott, Abbott,
  {et~al.}}]{collaboration2019gravitationalwave}
Abbott, B.~P., Abbott, R., Abbott, T.~D., {et~al.} 2019{\natexlab{a}},
  arXiv:1908.06060

\bibitem[{Abbott {et~al.}(2016{\natexlab{b}})Abbott, Abbott, Abbott, Abernathy,
  Acernese, Ackley, Adamo, Adams, Adams, Addesso, Adhikari, Adya, Affeldt,
  Agathos, {et~al.}}]{Abbott_2016}
---. 2016{\natexlab{b}}, Classical and Quantum Gravity, 33, 134001

\bibitem[{Abbott {et~al.}(2016{\natexlab{c}})Abbott, Abbott, Abbott, Abernathy,
  Acernese, Ackley, Adams, Adams, Addesso, Adhikari, Adya, Affeldt, Agathos,
  Agatsuma, Aggarwal, Aguiar, Aiello, Ain, Ajith, Allen, Allocca, Altin,
  Anderson, Anderson, Arai, Arain, Araya, Arceneaux, Areeda, Arnaud, Arun,
  Ascenzi, Ashton, Ast, Aston, Astone, Aufmuth, Aulbert, Babak, Bacon, Bader,
  Baker, \& et. al}]{PhysRevLett.116.061102}
---. 2016{\natexlab{c}}, Phys. Rev. Lett., 116, 061102

\bibitem[{{Abbott} {et~al.}(2017){Abbott}, {Abbott}, {Abbott}, {Acernese},
  {Ackley}, {Adams}, {Adams}, {Addesso}, {Adhikari}, {Adya}, {Affeldt}, \&
  {Agarwal}}]{2017Natur.551...85A}
{Abbott}, B.~P., {Abbott}, R., {Abbott}, T.~D., {et~al.} 2017, Nature, 551, 85

\bibitem[{Abbott {et~al.}(2017{\natexlab{a}})Abbott, Abbott, Abbott, Acernese,
  Ackley, Adams, Adams, Addesso, Adhikari, Adya,
  {et~al.}}]{abbott2017gravitational}
Abbott, B.~P., Abbott, R., Abbott, T., {et~al.} 2017{\natexlab{a}}, The
  Astrophysical Journal Letters, 848, L13

\bibitem[{Abbott {et~al.}(2017{\natexlab{b}})Abbott, Abbott, Abbott, Acernese,
  Ackley, Adams, Adams, Addesso, Adhikari, Adya, Affeldt, Afrough, Agarwal,
  Agathos, {et~al.}}]{PhysRevLett.119.161101}
Abbott, B.~P., Abbott, R., Abbott, T.~D., {et~al.} 2017{\natexlab{b}}, Phys.
  Rev. Lett., 119, 161101

\bibitem[{Abbott {et~al.}(2017{\natexlab{c}})Abbott, Abbott, Abbott, Acernese,
  Ackley, Adams, Adams, Addesso, Adhikari, Adya, Affeldt, Afrough, Agarwal,
  Agathos, Agatsuma, Aggarwal, Aguiar, Aiello, Ain, Ajith, Allen, Allen, \&
  A.~Allocca}]{Abbott_2017}
---. 2017{\natexlab{c}}, The Astrophysical Journal, 848, L12

\bibitem[{Abbott {et~al.}(2017{\natexlab{d}})Abbott, Abbott, Abbott, Abernathy,
  Acernese, Ackley, Adams, Adams, Addesso, Adhikari, Adya, Affeldt, Agathos,
  Agatsuma, Aggarwal, Aguiar, Aiello, Ain, Ajith, Allen, Allocca, Altin,
  Ananyeva, Anderson, Anderson, Appert, Arai, Araya, Areeda, \&
  and}]{Abbott_2017_150906B}
---. 2017{\natexlab{d}}, The Astrophysical Journal, 841, 89

\bibitem[{Abbott {et~al.}(2019{\natexlab{b}})Abbott, Abbott, Abbott, Abraham,
  Acernese, Ackley, Adams, Adhikari, Adya, Affeldt, Agathos, Agatsuma,
  Aggarwal, Aguiar, Aiello, Ain, Ajith, Allen, Allocca, Aloy, Altin, Amato,
  Anand, Ananyeva, Anderson, Anderson, Angelova, Antier, Appert, Arai, Araya,
  Areeda, Ar{\`{e}}ne, Arnaud, Aronson, Ascenzi, Ashton, Aston, Astone, Aubin,
  Aufmuth, AultONeal, Austin, Avendano, Avila, {et~al.}}]{Abbott_2019}
---. 2019{\natexlab{b}}, The Astrophysical Journal, 886, 75

\bibitem[{Abbott {et~al.}(2019{\natexlab{c}})Abbott, Abbott, Abbott, Abraham,
  Acernese, Ackley, Adams, Adhikari, Adya, Affeldt, Agathos, Agatsuma,
  Aggarwal, Aguiar, Aiello, Ain, Ajith, Allen, Allocca, Aloy, Altin, Amato,
  Ananyeva, Anderson, Anderson, Angelova, Antier, Appert, Arai, Araya, Areeda,
  Ar\`ene, Arnaud, Arun, Ascenzi, Ashton, Aston, Astone, Aubin, Aufmuth,
  AultONeal, Austin, Avendano, Avila-Alvarez, Babak, Bacon, Badaracco, Bader,
  Bae, Baker, Baldaccini, Ballardin, Ballmer, Banagiri, Barayoga, Barclay,
  Barish, Barker, Barkett, Barnum, Barone, Barr, Barsotti, Barsuglia, Barta,
  Bartlett, Bartos, Bassiri, Basti, Bawaj, Bayley, Bazzan, B\'ecsy, Bejger,
  Belahcene, Bell, Beniwal, Berger, Bergmann, Bernuzzi, Bero, Berry,
  Bersanetti, Bertolini, Betzwieser, Bhandare, Bidler, Bilenko, Bilgili,
  Billingsley, Birch, Birney, Birnholtz, Biscans, Biscoveanu, Bisht, Bitossi,
  Bizouard, Blackburn, Blackman, Blair, Blair, Blair, Bloemen, Bode, Boer,
  Boetzel, Bogaert, Bondu, Bonilla, Bonnand, Booker, Boom, Booth, Bork, \&
  Boschi}]{2018arXiv181112907T}
---. 2019{\natexlab{c}}, Phys. Rev. X, 9, 031040

\bibitem[{Abbott {et~al.}(2019{\natexlab{d}})Abbott, Abbott, Abbott, Acernese,
  Ackley, Adams, Adams, Addesso, Adhikari, Adya, Affeldt, \&
  Agarwal}]{PhysRevX.9.011001}
---. 2019{\natexlab{d}}, Phys. Rev. X, 9, 011001

\bibitem[{Abbott {et~al.}(2019{\natexlab{e}})Abbott, Abbott, Abbott, Abraham,
  Acernese, Ackley, Adams, Adya, Affeldt, Agathos, Agatsuma, Aggarwal, Aguiar,
  Aiello, Ain, Ajith, Akutsu, Allen, Allocca, Aloy, Altin, Amato, Ananyeva,
  Anderson, Anderson, Ando, Angelova, Antier, Appert, Arai, Arai, Arai, Araki,
  Araya, Araya, Areeda, Arène, Aritomi, Arnaud, Arun, Ascenzi, Ashton, Aso,
  Aston, Astone, Aubin, Aufmuth, AultONeal, Austin, Avendano, \&
  and}]{collaboration2013prospects}
---. 2019{\natexlab{e}}, arXiv:1304.0670

\bibitem[{Abbott {et~al.}(2019{\natexlab{f}})Abbott, Abbott, Abbott, Abraham,
  Acernese, Ackley, Adams, Adams, Adhikari, Adya, Affeldt, Agathos, Agatsuma,
  Aggarwal, Aguiar, Aiello, Ain, Ajith, Allen, Allocca, Aloy, Altin, Amato,
  Anand, Ananyeva, Anderson, Anderson, Angelova, Antier, Appert, Arai, Araya,
  Areeda, Ar\`ene, Arnaud, Aronson, Arun, Ascenzi, Ashton, Aston, Astone,
  Aubin, Aufmuth, AultONeal, Austin, Avendano, \&
  Avila-Alvarez}]{PhysRevD.100.064064}
---. 2019{\natexlab{f}}, Phys. Rev. D, 100, 064064

\bibitem[{Acernese {et~al.}(2015)}]{TheVirgo:2014hva}
Acernese, F., {et~al.} 2015, Class. Quant. Grav., 32, 024001

\bibitem[{Ashton {et~al.}(2018)Ashton, Burns, {Dal Canton}, Dent, Eggenstein,
  Nielsen, Prix, Was, \& Zhu}]{Ashton_2018}
Ashton, G., Burns, E., {Dal Canton}, T., {et~al.} 2018, The Astrophysical
  Journal, 860, 6

\bibitem[{{Astropy Collaboration} {et~al.}(2013){Astropy Collaboration},
  {Robitaille}, {Tollerud}, {Greenfield}, {Droettboom}, {Bray}, {Aldcroft},
  {Davis}, {Ginsburg}, {Price-Whelan}, {Kerzendorf}, {Conley}, {Crighton},
  {Barbary}, {Muna}, {Ferguson}, {Grollier}, {Parikh}, {Nair}, {Unther},
  {Deil}, {Woillez}, {Conseil}, {Kramer}, {Turner}, {Singer}, {Fox}, {Weaver},
  {Zabalza}, {Edwards}, {Azalee Bostroem}, {Burke}, {Casey}, {Crawford},
  {Dencheva}, {Ely}, {Jenness}, {Labrie}, {Lim}, {Pierfederici}, {Pontzen},
  {Ptak}, {Refsdal}, {Servillat}, \& {Streicher}}]{2013A&A...558A..33A}
{Astropy Collaboration}, {Robitaille}, T.~P., {Tollerud}, E.~J., {et~al.} 2013,
  \aap, 558, A33

\bibitem[{Bhat {et~al.}(2016)Bhat, Meegan, von Kienlin, Paciesas, Briggs,
  Burgess, Burns, Chaplin, Cleveland, Collazzi, Connaughton, Diekmann,
  Fitzpatrick, Gibby, Giles, Goldstein, Greiner, Jenke, Kippen, Kouveliotou,
  Mailyan, McBreen, Pelassa, Preece, Roberts, Sparke, Stanbro, Veres,
  Wilson-Hodge, Xiong, Younes, Yu, \& Zhang}]{GBMBurstCatalog_6Years}
Bhat, P.~N., Meegan, C.~A., von Kienlin, A., {et~al.} 2016, ApJS, 223, 28

\bibitem[{{Blackburn} {et~al.}(2015){Blackburn}, {Briggs}, {Camp},
  {Christensen}, {Connaughton}, {Jenke}, {Remillard}, \&
  {Veitch}}]{Blackburn2015}
{Blackburn}, L., {Briggs}, M.~S., {Camp}, J., {et~al.} 2015, \apjs, 217, 8

\bibitem[{Burns {et~al.}(2019)Burns, Goldstein, Hui, Blackburn, Briggs,
  Connaughton, Hamburg, Kocevski, Veres, Wilson-Hodge, Bissaldi, Cleveland,
  Giles, Mailyan, Meegan, Paciesas, Poolakkil, Preece, Racusin, Roberts, von
  Kienlin, Abbott, Abbott, Abbott, Acernese, Ackley, Adams, Adams, Addesso,
  Adhikari, Adya, Affeldt, Agarwal, Agathos, Agatsuma, Aggarwal, Aguiar,
  Aiello, Ain, Ajith, Allen, Allen, Allocca, Aloy, Altin, Amato, Ananyeva,
  Anderson, Anderson, Angelova, Antier, Appert, Arai, Araya, Areeda,
  Ar{\`{e}}ne, Arnaud, Arun, Ascenzi, Ashton, Ast, Aston, Astone, Atallah,
  Aubin, Aufmuth, Aulbert, AultONeal, Austin, Avila-Alvarez, Babak, Bacon,
  Badaracco, Bader, Bae, Baker, Baldaccini, Ballardin, Ballmer, Banagiri,
  Barayoga, Barclay, Barish, Barker, Barkett, Barnum, Barone, Barr, Barsotti,
  Barsuglia, Barta, Bartlett, Bartos, Bassiri, Basti, Batch, Bawaj, Bayley,
  Bazzan, B{\'{e}}csy, Beer, Bejger, Belahcene, Bell, Beniwal, Bensch, Berger,
  Bergmann, Bernuzzi, Bero, Berry, Bersanetti, Bertolini, Betzwieser, Bhandare,
  Bilenko, Bilgili, Billingsley, Billman, Birch, Birney, Birnholtz, Biscans,
  Biscoveanu, Bisht, Bitossi, Bizouard, Blackburn, Blackman, Blair, Blair,
  Blair, Bloemen, Bock, Bode, Boer, Boetzel, Bogaert, Bohe, Bondu, Bonilla,
  Bonnand, Booker, Boom, Booth, Bork, Boschi, Bose, Bossie, Bossilkov, Bosveld,
  Bouffanais, Bozzi, Bradaschia, Brady, Bramley, Branchesi, Brau, Briant,
  Brighenti, Brillet, Brinkmann, Brisson, Brockill, Brooks, Brown, Brunett,
  Buchanan, Buikema, Bulik, Bulten, Buonanno, Buskulic, Buy, Byer, Cabero,
  Cadonati, Cagnoli, Cahillane, Bustillo, Callister, Calloni, Camp, Canepa,
  Canizares, Cannon, Cao, Cao, Capano, Capocasa, Carbognani, Caride, Carney,
  Diaz, Casentini, Caudill, Cavagli{\`{a}}, Cavalier, Cavalieri, Cella, Cepeda,
  Cerd{\'{a}}-Dur{\'{a}}n, Cerretani, Cesarini, Chaibi, Chamberlin, Chan, Chao,
  Charlton, Chase, Chassande-Mottin, Chatterjee, Cheeseboro, Chen, Chen, Chen,
  Cheng, Chia, Chincarini, Chiummo, Chmiel, Cho, Cho, Chow, Christensen, Chu,
  Chua, Chua, Chung, Chung, Ciani, Ciobanu, Ciolfi, Cipriano, Cirelli, Cirone,
  Clara, Clark, Clearwater, Cleva, Cocchieri, Coccia, Cohadon, Cohen, Colla,
  Collette, Collins, Cominsky, Jr., Conti, Cooper, Corban, Corbitt,
  Cordero-Carri{\'{o}}n, Corley, Cornish, Corsi, Cortese, Costa, Cotesta,
  Coughlin, Coughlin, Coulon, Countryman, Couvares, Covas, Cowan, Coward,
  Cowart, Coyne, Coyne, Creighton, Creighton, Cripe, Crowder, Cullen, Cumming,
  Cunningham, Cuoco, Canton, D{\'{a}}lya, Danilishin, D'Antonio, Danzmann,
  Dasgupta, Costa, Dattilo, Dave, Davier, Davis, Daw, Day, DeBra, Deenadayalan,
  Degallaix, Laurentis, Del{\'{e}}glise, Pozzo, Demos, Denker, Dent, Pietri,
  Derby, Dergachev, Rosa, Rossi, DeSalvo, de~Varona, Dhurandhar, D{\'{\i}}az,
  Fiore, Giovanni, Girolamo, Lieto, Ding, Pace, Palma, Renzo, Dmitriev, Doctor,
  Dolique, Donovan, Dooley, Doravari, Dorrington, {\'{A}}lvarez, Downes, Drago,
  Dreissigacker, Driggers, Du, Dupej, Dwyer, Easter, Edo, Edwards, Effler,
  Eggenstein, Ehrens, Eichholz, Eikenberry, Eisenmann, Eisenstein, Essick,
  Estelles, Estevez, Etienne, Etzel, Evans, Evans, Fafone, Fair, Fairhurst,
  Fan, Farinon, Farr, Farr, Fauchon-Jones, Favata, Fays, Fee, Fehrmann, Feicht,
  Fejer, Feng, Fernandez-Galiana, Ferrante, Ferreira, Ferrini, Fidecaro, Fiori,
  Fiorucci, Fishbach, Fisher, Fishner, Fitz-Axen, Flaminio, Fletcher, Fong,
  Font, Forsyth, Forsyth, Fournier, Frasca, Frasconi, Frei, Freise, Frey, Frey,
  Fritschel, Frolov, Fulda, Fyffe, Gabbard, Gadre, Gaebel, Gair, Gammaitoni,
  Ganija, Gaonkar, Garcia, Garc{\'{\i}}a-Quir{\'{o}}s, Garufi, Gateley, Gaudio,
  Gaur, Gayathri, Gemme, Genin, Gennai, George, George, Gergely, Germain,
  Ghonge, Ghosh, Ghosh, Ghosh, Giacomazzo, Giaime, Giardina, Giazotto, Gill,
  Giordano, Glover, Goetz, Goetz, Goncharov, Gonz{\'{a}}lez, Castro, Gopakumar,
  Gorodetsky, Gossan, Gosselin, Gouaty, Grado, Graef, Granata, Grant, Gras,
  Gray, Greco, Green, Green, Gretarsson, Groot, Grote, Grunewald, Gruning,
  Guidi, Gulati, Guo, Gupta, Gupta, Gushwa, Gustafson, Gustafson, Halim, Hall,
  Hall, Hamilton, Hamilton, Hammond, Haney, Hanke, Hanks, Hanna, Hannam,
  Hannuksela, Hanson, Hardwick, Harms, Harry, Harry, Hart, Haster, Haughian,
  Healy, Heidmann, Heintze, Heitmann, Hello, Hemming, Hendry, Heng, Hennig,
  Heptonstall, Hernandez, Heurs, Hild, Hinderer, Hoak, Hochheim, Hofman,
  Holland, Holt, Holz, Hopkins, Horst, Hough, Houston, Howell, Hreibi, Huerta,
  Huet, Hughey, Hulko, Husa, Huttner, Huynh-Dinh, Iess, Indik, Ingram, Inta,
  Intini, Isa, Isac, Isi, Iyer, Izumi, Jacqmin, Jani, Jaranowski, Johnson,
  Johnson, Jones, Jones, Jonker, Ju, Junker, Kalaghatgi, Kalogera, Kamai,
  Kandhasamy, Kang, Kanner, Kapadia, Karki, Karvinen, Kasprzack, Katolik,
  Katsanevas, Katsavounidis, Katzman, Kaufer, Kawabe, Keerthana,
  K{\'{e}}f{\'{e}}lian, Keitel, Kemball, Kennedy, Key, Khalili, Khamesra, Khan,
  Khan, Khan, Khan, Khazanov, Kijbunchoo, Kim, Kim, Kim, Kim, Kim, Kim, King,
  King, Kinley-Hanlon, Kirchhoff, Kissel, Kleybolte, Klimenko, Knowles, Koch,
  Koehlenbeck, Koley, Kondrashov, Kontos, Korobko, Korth, Kowalska, Kozak,
  Krämer, Kringel, Krishnan, Kr{\'{o}}lak, Kuehn, Kumar, Kumar, Kumar, Kuo,
  Kutynia, Kwang, Lackey, Lai, Landry, Lang, Lange, Lantz, Lanza,
  Lartaux-Vollard, Lasky, Laxen, Lazzarini, Lazzaro, Leaci, Leavey, Lee, Lee,
  Lee, Lee, Lee, Lehmann, Lenon, Leonardi, Leroy, Letendre, Levin, Li, Li, Li,
  Linker, Littenberg, Liu, Liu, Lo, Lockerbie, London, Longo, Lorenzini,
  Loriette, Lormand, Losurdo, Lough, Lousto, Lovelace, Lück, Lumaca, Lundgren,
  Lynch, Ma, Macas, Macfoy, Machenschalk, MacInnis, Macleod, Hernandez,
  Maga{\~{n}}a-Sandoval, Zertuche, Magee, Majorana, Maksimovic, Man, Mandic,
  Mangano, Mansell, Manske, Mantovani, Marchesoni, Marion, M{\'{a}}rka,
  M{\'{a}}rka, Markakis, Markosyan, Markowitz, Maros, Marquina, Martelli,
  Martellini, Martin, Martin, Martynov, Mason, Massera, Masserot, Massinger,
  Masso-Reid, Mastrogiovanni, Matas, Matichard, Matone, Mavalvala, Mazumder,
  McCann, McCarthy, McClelland, McCormick, McCuller, McGuire, McIver, McManus,
  McRae, McWilliams, Meacher, Meadors, Mehmet, Meidam, Mejuto-Villa, Melatos,
  Mendell, Mendoza-Gandara, Mercer, Mereni, Merilh, Merzougui, Meshkov,
  Messenger, Messick, Metzdorff, Meyers, Miao, Michel, Middleton, Mikhailov,
  Milano, Miller, Miller, Miller, Miller, Millhouse, Mills, Milovich-Goff,
  Minazzoli, Minenkov, Ming, Mishra, Mitra, Mitrofanov, Mitselmakher,
  Mittleman, Moffa, Mogushi, Mohan, Mohapatra, Montani, Moore, Moraru, Moreno,
  Morisaki, Mours, Mow-Lowry, Mueller, Muir, Mukherjee, Mukherjee, Mukherjee,
  Mukund, Mullavey, Munch, Mu{\~{n}}iz, Muratore, Murray, Nagar, Napier,
  Nardecchia, Naticchioni, Nayak, Neilson, Nelemans, Nelson, Nery, Neunzert,
  Nevin, Newport, Ng, Ng, Nguyen, Nguyen, Nichols, Nielsen, Nissanke, Nitz,
  Nocera, Nolting, North, Nuttall, Obergaulinger, Oberling, O'Brien, O'Dea,
  Ogin, Oh, Oh, Ohme, Ohta, Okada, Oliver, Oppermann, Oram, O'Reilly, Ormiston,
  Ortega, O'Shaughnessy, Ossokine, Ottaway, Overmier, Owen, Pace, Pagano, Page,
  Page, Pai, Pai, Palamos, Palashov, Palomba, Pal-Singh, Pan, Pan, Pang, Pang,
  Pankow, Pannarale, Pant, Paoletti, Paoli, Papa, Parida, Parker, Pascucci,
  Pasqualetti, Passaquieti, Passuello, Patil, Patricelli, Pearlstone, Pedersen,
  Pedraza, Pedurand, Pekowsky, Pele, Penn, Perez, Perreca, Perri, Pfeiffer,
  Phelps, Phukon, Piccinni, Pichot, Piergiovanni, Pierro, Pillant, Pinard,
  Pinto, Pirello, Pitkin, Poggiani, Popolizio, Porter, Possenti, Post, Powell,
  Prasad, Pratt, Pratten, Predoi, Prestegard, Principe, Privitera, Prodi,
  Prokhorov, Puncken, Punturo, Puppo, Pürrer, Qi, Quetschke, Quintero,
  Quitzow-James, Raab, Rabeling, Radkins, Raffai, Raja, Rajan, Rajbhandari,
  Rakhmanov, Ramirez, Ramos-Buades, Rana, Rapagnani, Raymond, Razzano, Read,
  Regimbau, Rei, Reid, Reitze, Ren, Ricci, Ricker, Riles, Rizzo, Robertson,
  Robie, Robinet, Robson, Rocchi, Rolland, Rollins, Roma, Romano, Romel, Romie,
  Rosi{\'{n}}ska, Ross, Rowan, Rüdiger, Ruggi, Rutins, Ryan, Sachdev, Sadecki,
  Sakellariadou, Salconi, Saleem, Salemi, Samajdar, Sammut, Sampson, Sanchez,
  Sanchez, Sanchis-Gual, Sandberg, Sanders, Sarin, Sassolas, Sathyaprakash,
  Saulson, Sauter, Savage, Sawadsky, Schale, Scheel, Scheuer, Schmidt,
  Schnabel, Schofield, Schönbeck, Schreiber, Schuette, Schulte, Schutz,
  Schwalbe, Scott, Scott, Seidel, Sellers, Sengupta, Sentenac, Sequino,
  Sergeev, Setyawati, Shaddock, Shaffer, Shah, Shahriar, Shaner, Shao, Shapiro,
  Shawhan, Shen, Shoemaker, Shoemaker, Siellez, Siemens, Sieniawska, Sigg,
  Silva, Singer, Singh, Singhal, Sintes, Slagmolen, Slaven-Blair, Smith, Smith,
  Smith, Somala, Son, Sorazu, Sorrentino, Souradeep, Spencer, Srivastava,
  Staats, Stachie, Steinke, Steinlechner, Steinlechner, Steinmeyer, Steltner,
  Stevenson, Stocks, Stone, Stops, Strain, Stratta, Strigin, Strunk, Sturani,
  Stuver, Summerscales, Sun, Sunil, Suresh, Sutton, Swinkels,
  Szczepa{\'{n}}czyk, Tacca, Tait, Talbot, Talukder, Tanner, T{\'{a}}pai,
  Taracchini, Tasson, Taylor, Taylor, Tewari, Theeg, Thies, Thomas, Thomas,
  Thomas, Thorne, Thrane, Tiwari, Tiwari, Tokmakov, Toland, Tonelli, Tornasi,
  Torres-Forn{\'{e}}, Torrie, Töyrä, Travasso, Traylor, Trinastic, Tringali,
  Trozzo, Tsang, Tse, Tso, Tsuna, Tsukada, Tuyenbayev, Ueno, Ugolini, Urban,
  Usman, Vahlbruch, Vajente, Valdes, van Bakel, van Beuzekom, van~den Brand,
  Broeck, Vander-Hyde, van~der Schaaf, van Heijningen, van Veggel, Vardaro,
  Varma, Vass, Vas{\'{u}}th, Vecchio, Vedovato, Veitch, Veitch, Venkateswara,
  Venugopalan, Verkindt, Vetrano, Vicer{\'{e}}, Viets, Vinciguerra, Vine,
  Vinet, Vitale, Vo, Vocca, Vorvick, Vyatchanin, Wade, Wade, Wade, Walet,
  Walker, Wallace, Walsh, Wang, Wang, Wang, Wang, Wang, Ward, Warner, Was,
  Watchi, Weaver, Wei, Weinert, Weinstein, Weiss, Wellmann, Wen, Wessel,
  We{\ss}els, Westerweck, Wette, Whelan, Whiting, Whittle, Wilken, Williams,
  Williams, Williamson, Willis, Willke, Wimmer, Winkler, Wipf, Wittel, Woan,
  Woehler, Wofford, Wong, Worden, Wright, Wu, Wysocki, Xiao, Yam, Yamamoto,
  Yancey, Yang, Yap, Yazback, Yu, Yu, Yvert, Zadro{\.{z}}ny, Zanolin, Zelenova,
  Zendri, Zevin, Zhang, Zhang, Zhang, Zhang, Zhang, Zhao, Zhou, Zhou, Zhu, Zhu,
  Zucker, Zweizig, \& and}]{Burns_2019}
Burns, E., Goldstein, A., Hui, C.~M., {et~al.} 2019, The Astrophysical Journal,
  871, 90

\bibitem[{Cabero {et~al.}(2019)Cabero, Lundgren, Nitz, Dent, Barker, Goetz,
  Kissel, Nuttall, Schale, Schofield, {et~al.}}]{cabero2019blip}
Cabero, M., Lundgren, A., Nitz, A.~H., {et~al.} 2019, arXiv preprint
  arXiv:1901.05093

\bibitem[{Callister {et~al.}(2017)Callister, Kanner, Massinger, Dhurandhar, \&
  Weinstein}]{Callister:2017urp}
Callister, T.~A., Kanner, J.~B., Massinger, T.~J., Dhurandhar, S., \&
  Weinstein, A.~J. 2017, Class. Quant. Grav., 34, 155007

\bibitem[{Chornock {et~al.}(2017)Chornock, Berger, Kasen, Cowperthwaite,
  Nicholl, Villar, Alexander, Blanchard, Eftekhari, Fong, Margutti, Williams,
  Annis, Brout, Brown, Chen, Drout, Farr, Foley, Frieman, Fryer, Herner, Holz,
  Kessler, Matheson, Metzger, Quataert, Rest, Sako, Scolnic, Smith, \&
  Soares-Santos}]{Chornock_2017}
Chornock, R., Berger, E., Kasen, D., {et~al.} 2017, The Astrophysical Journal,
  848, L19

\bibitem[{Connaughton {et~al.}(2015)Connaughton, Briggs, Goldstein, Meegan,
  Paciesas, Preece, Wilson-Hodge, Gibby, Greiner, Gruber, Jenke, Kippen,
  Pelassa, Xiong, Yu, Bhat, Burgess, Byrne, Fitzpatrick, Foley, Giles, Guiriec,
  van~der Horst, von Kienlin, McBreen, McGlynn, Tierney, \&
  Zhang}]{Connaughton_2015}
Connaughton, V., Briggs, M.~S., Goldstein, A., {et~al.} 2015, The Astrophysical
  Journal Supplement Series, 216, 32

\bibitem[{Connaughton {et~al.}(2016)Connaughton, Burns, Goldstein, Blackburn,
  Briggs, Zhang, Camp, Christensen, Hui, Jenke, Littenberg, McEnery, Racusin,
  Shawhan, Singer, Veitch, Wilson-Hodge, Bhat, Bissaldi, Cleveland,
  Fitzpatrick, Giles, Gibby, von Kienlin, Kippen, McBreen, Mailyan, Meegan,
  Paciesas, Preece, Roberts, Sparke, Stanbro, Toelge, \&
  Veres}]{Connaughton_2016}
Connaughton, V., Burns, E., Goldstein, A., {et~al.} 2016, The Astrophysical
  Journal, 826, L6

\bibitem[{Connaughton {et~al.}(2018)Connaughton, Burns, Goldstein, Blackburn,
  Briggs, Christensen, Hui, Kocevski, Littenberg, McEnery, Racusin, Shawhan,
  Veitch, Wilson-Hodge, Bhat, Bissaldi, Cleveland, Giles, Gibby, von Kienlin,
  Kippen, McBreen, Meegan, Paciesas, Preece, Roberts, Stanbro, \&
  Veres}]{Connaughton_2018}
---. 2018, The Astrophysical Journal, 853, L9

\bibitem[{Cowperthwaite {et~al.}(2017)Cowperthwaite, Berger, Villar, Metzger,
  Nicholl, Chornock, Blanchard, Fong, Margutti, Soares-Santos, Alexander,
  Allam, Annis, Brout, Brown, Butler, Chen, Diehl, Doctor, Drout, Eftekhari,
  Farr, Finley, Foley, Frieman, Fryer, Garc{\'{\i}}a-Bellido, Gill, Guillochon,
  Herner, Holz, Kasen, Kessler, Marriner, Matheson, Neilsen, Quataert, Palmese,
  Rest, Sako, Scolnic, Smith, Tucker, Williams, Balbinot, Carlin, Cook, Durret,
  Li, Lopes, Louren{\c{c}}o, Marshall, Medina, Muir, Mu{\~{n}}oz, Sauseda,
  Schlegel, Secco, \& and}]{Cowperthwaite_2017}
Cowperthwaite, P.~S., Berger, E., Villar, V.~A., {et~al.} 2017, The
  Astrophysical Journal, 848, L17

\bibitem[{Dal~Canton \& Harry(2017)}]{DalCanton:2017ala}
Dal~Canton, T., \& Harry, I.~W. 2017, arXiv:1705.01845

\bibitem[{Goldstein {et~al.}(2016)Goldstein, Burns, Hamburg, Connaughton,
  Veres, Briggs, Hui, \& Collaboration}]{goldstein2016updates}
Goldstein, A., Burns, E., Hamburg, R., {et~al.} 2016, arXiv preprint
  arXiv:1612.02395

\bibitem[{Goldstein {et~al.}(2017)Goldstein, Veres, Burns, Briggs, Hamburg,
  Kocevski, Wilson-Hodge, Preece, Poolakkil, Roberts, Hui, Connaughton,
  Racusin, von Kienlin, Canton, Christensen, Littenberg, Siellez, Blackburn,
  Broida, Bissaldi, Cleveland, Gibby, Giles, Kippen, McBreen, McEnery, Meegan,
  Paciesas, \& Stanbro}]{Goldstein_2017b}
Goldstein, A., Veres, P., Burns, E., {et~al.} 2017, The Astrophysical Journal,
  848, L14

\bibitem[{Goldstein {et~al.}(2019)Goldstein, Hamburg, Michelle~Hui,
  H.~Cleveland, Kocevski, Littenberg, Burns, Dal~Canton, Veres, Mailyan,
  Malacaria, S.~Briggs, \& A.~Wilson-Hodge}]{goldstein2019updates}
Goldstein, A., Hamburg, R., Michelle~Hui, C., {et~al.} 2019, arXiv preprint
  arXiv:1903.12597

\bibitem[{{G{\'o}rski} {et~al.}(2005){G{\'o}rski}, {Hivon}, {Banday},
  {Wandelt}, {Hansen}, {Reinecke}, \& {Bartelmann}}]{2005ApJ...622..759G}
{G{\'o}rski}, K.~M., {Hivon}, E., {Banday}, A.~J., {et~al.} 2005, \apj, 622,
  759

\bibitem[{Greiner {et~al.}(2016)Greiner, Burgess, Savchenko, \&
  Yu}]{Greiner_2016}
Greiner, J., Burgess, J.~M., Savchenko, V., \& Yu, H.-F. 2016, The
  Astrophysical Journal, 827, L38

\bibitem[{{Hotokezaka} {et~al.}(2019){Hotokezaka}, {Nakar}, {Gottlieb},
  {Nissanke}, {Masuda}, {Hallinan}, {Mooley}, \&
  {Deller}}]{2019NatAs.tmp..385H}
{Hotokezaka}, K., {Nakar}, E., {Gottlieb}, O., {et~al.} 2019, Nature Astronomy,
  385

\bibitem[{Hunter(2007)}]{Hunter:2007}
Hunter, J.~D. 2007, Computing In Science \& Engineering, 9, 90

\bibitem[{Isogai {et~al.}(2010)Isogai, the Ligo Scientific~Collaboration, \&
  the Virgo~Collaboration}]{Isogai_2010}
Isogai, T., the Ligo Scientific~Collaboration, \& the Virgo~Collaboration.
  2010, Journal of Physics: Conference Series, 243, 012005

\bibitem[{Jones {et~al.}(2001)Jones, Oliphant, Peterson, \&
  Others}]{jones_scipy_2001}
Jones, E., Oliphant, T., Peterson, P., \& Others. 2001, {SciPy}: Open source
  scientific tools for Python, ,

\bibitem[{{Kapadia} {et~al.}(2019){Kapadia}, {Caudill}, {Creighton}, {Farr},
  {Mendell}, {Weinstein}, {Cannon}, {Fong}, {Godwin}, {Lo}, {Magee}, {Meacher},
  {Messick}, {Mohite}, {Mukherjee}, \& {Sachdev}}]{2019arXiv190306881K}
{Kapadia}, S.~J., {Caudill}, S., {Creighton}, J. D.~E., {et~al.} 2019, arXiv
  e-prints, arXiv:1903.06881

\bibitem[{{Kasen} {et~al.}(2017){Kasen}, {Metzger}, {Barnes}, {Quataert}, \&
  {Ramirez-Ruiz}}]{2017Natur.551...80K}
{Kasen}, D., {Metzger}, B., {Barnes}, J., {Quataert}, E., \& {Ramirez-Ruiz}, E.
  2017, \nat, 551, 80

\bibitem[{{LIGO Scientific \& Virgo Collaboration}(2018)}]{GWTC1_skymaps}
{LIGO Scientific \& Virgo Collaboration}. 2018, Sky localization probability
  maps (skymaps) release for GWTC-1,
  \url{https://dcc.ligo.org/LIGO-P1800381/public}, ,

\bibitem[{Meegan {et~al.}(2009)Meegan, Lichti, Bhat, Bissaldi, Briggs,
  Connaughton, Diehl, Fishman, Greiner, Hoover, van~der Horst, von Kienlin,
  Kippen, Kouveliotou, McBreen, Paciesas, Preece, Steinle, Wallace, Wilson, \&
  Wilson-Hodge}]{Meegan_2009}
Meegan, C., Lichti, G., Bhat, P.~N., {et~al.} 2009, The Astrophysical Journal,
  702, 791

\bibitem[{Messick {et~al.}(2017)Messick, Blackburn, Brady, Brockill, Cannon,
  Cariou, Caudill, Chamberlin, Creighton, Everett, Hanna, Keppel, Lang, Li,
  Meacher, Nielsen, Pankow, Privitera, Qi, Sachdev, Sadeghian, Singer, Thomas,
  Wade, Wade, Weinstein, \& Wiesner}]{PhysRevD.95.042001}
Messick, C., Blackburn, K., Brady, P., {et~al.} 2017, Phys. Rev. D, 95, 042001

\bibitem[{{Mukherjee} {et~al.}(2018){Mukherjee}, {Caudill}, {Magee}, {Messick},
  {Privitera}, {Sachdev}, {Blackburn}, {Brady}, {Brockill}, {Cannon},
  {Chamberlin}, {Chatterjee}, {Creighton}, {Fong}, {Godwin}, {Hanna},
  {Kapadia}, {Lang}, {Li}, {Lo}, {Meacher}, {Pace}, {Sadeghian}, {Tsukada},
  {Wade}, {Wade}, {Weinstein}, \& {Xiao}}]{2018arXiv181205121M}
{Mukherjee}, D., {Caudill}, S., {Magee}, R., {et~al.} 2018, arXiv e-prints,
  arXiv:1812.05121

\bibitem[{Nitz {et~al.}(2017)Nitz, Dent, {Dal Canton}, Fairhurst, \&
  Brown}]{Nitz:2017}
Nitz, A.~H., Dent, T., {Dal Canton}, T., Fairhurst, S., \& Brown, D.~A. 2017,
  The Astrophysical Journal, 849, 118

\bibitem[{Nitz {et~al.}(2019)Nitz, Nielsen, \& Capano}]{Nitz_2019}
Nitz, A.~H., Nielsen, A.~B., \& Capano, C.~D. 2019, The Astrophysical Journal,
  876, L4

\bibitem[{{Pankow} {et~al.}(2018){Pankow}, {Chatziioannou}, {Chase},
  {Littenberg}, {Evans}, {McIver}, {Cornish}, {Haster}, {Kanner}, {Raymond},
  {Vitale}, \& {Zimmerman}}]{2018PhRvD..98h4016P}
{Pankow}, C., {Chatziioannou}, K., {Chase}, E.~A., {et~al.} 2018, Phys. Rev. D,
  98, 084016

\bibitem[{Sachdev {et~al.}(2019)}]{Sachdev:2019vvd}
Sachdev, S., {et~al.} 2019, arXiv:1901.08580

\bibitem[{Salafia {et~al.}(2018)Salafia, Ghisellini, Ghirlanda, \&
  Colpi}]{Salafia_2018}
Salafia, O.~S., Ghisellini, G., Ghirlanda, G., \& Colpi, M. 2018, Astronomy \&
  Astrophysics, 619, A18

\bibitem[{Savchenko {et~al.}(2017)Savchenko, Ferrigno, Kuulkers, Bazzano,
  Bozzo, Brandt, Chenevez, Courvoisier, Diehl, Domingo, Hanlon, Jourdain, von
  Kienlin, Laurent, Lebrun, Lutovinov, Martin-Carrillo, Mereghetti, Natalucci,
  Rodi, Roques, Sunyaev, \& Ubertini}]{Savchenko_2017}
Savchenko, V., Ferrigno, C., Kuulkers, E., {et~al.} 2017, The Astrophysical
  Journal, 848, L15

\bibitem[{Singer \& Price(2016)}]{PhysRevD.93.024013}
Singer, L.~P., \& Price, L.~R. 2016, Phys. Rev. D, 93, 024013

\bibitem[{Stachie {et~al.}(2020)Stachie, Canton, Burns, Christensen, Hamburg,
  Briggs, Broida, Goldstein, Hayes, Littenberg, Shawhan, Veitch, Veres, \&
  Wilson-Hodge}]{stachie2020search}
Stachie, C., Canton, T.~D., Burns, E., {et~al.} 2020, Search for Advanced LIGO
  Single Interferometer Compact Binary Coalescence Signals in Coincidence with
  Gamma-Ray Events in Fermi-GBM, , , arXiv:2001.01462

\bibitem[{Tanvir {et~al.}(2017)Tanvir, Levan, Gonz{\'{a}}lez-Fern{\'{a}}ndez,
  Korobkin, Mandel, Rosswog, Hjorth, D'Avanzo, Fruchter, Fryer, Kangas,
  Milvang-Jensen, Rosetti, Steeghs, Wollaeger, Cano, Copperwheat, Covino,
  D'Elia, de~Ugarte~Postigo, Evans, Even, Fairhurst, Jaimes, Fontes, Fujii,
  Fynbo, Gompertz, Greiner, Hodosan, Irwin, Jakobsson, J{\o}rgensen, Kann,
  Lyman, Malesani, McMahon, Melandri, O'Brien, Osborne, Palazzi, Perley, Pian,
  Piranomonte, Rabus, Rol, Rowlinson, Schulze, Sutton, Thöne, Ulaczyk, Watson,
  Wiersema, \& Wijers}]{Tanvir_2017}
Tanvir, N.~R., Levan, A.~J., Gonz{\'{a}}lez-Fern{\'{a}}ndez, C., {et~al.} 2017,
  The Astrophysical Journal, 848, L27

\bibitem[{Urban(2016)}]{UrbanThesis}
Urban, A.~L. 2016, PhD thesis, {University of Wisconsin Milwaukee}

\bibitem[{Usman {et~al.}(2016)Usman, Nitz, Harry, Biwer, Brown, Cabero, Capano,
  Canton, Dent, Fairhurst, Kehl, Keppel, Krishnan, Lenon, Lundgren, Nielsen,
  Pekowsky, Pfeiffer, Saulson, West, \& Willis}]{Usman_2016}
Usman, S.~A., Nitz, A.~H., Harry, I.~W., {et~al.} 2016, Classical and Quantum
  Gravity, 33, 215004

\bibitem[{Van Der~Walt {et~al.}(2011)Van Der~Walt, Colbert, \&
  Varoquaux}]{van2011numpy}
Van Der~Walt, S., Colbert, S.~C., \& Varoquaux, G. 2011, Computing in Science
  \& Engineering, 13, 22

\bibitem[{Vedrenne \& Atteia(2009)}]{Vedrenne2009}
Vedrenne, G., \& Atteia, J.-L. 2009, {Gamma-Ray Bursts: The Brightest
  Explosions in the Universe} (Berlin, Heidelberg: Springer Berlin Heidelberg),
  385--476

\bibitem[{Veitch \& Vecchio(2010)}]{PhysRevD.81.062003}
Veitch, J., \& Vecchio, A. 2010, Phys. Rev. D, 81, 062003

\bibitem[{Veitch {et~al.}(2015)Veitch, Raymond, Farr, Farr, Graff, Vitale,
  Aylott, Blackburn, Christensen, Coughlin, \& et~al.}]{Veitch_2015}
Veitch, J., Raymond, V., Farr, B., {et~al.} 2015, Physical Review D, 91,
  doi:10.1103/physrevd.91.042003

\bibitem[{Veres {et~al.}(2019)Veres, Dal~Canton, Burns, Goldstein, Littenberg,
  Christensen, Preece, \& D.}]{Veres:2019hsd}
Veres, P., Dal~Canton, T., Burns, E., {et~al.} 2019, The Astrophysical Journal,
  882, 53

\bibitem[{Watson {et~al.}(2019)Watson, Hansen, Selsing, Koch, Malesani,
  Andersen, Fynbo, Arcones, Bauswein, Covino, \& et~al.}]{Watson_2019}
Watson, D., Hansen, C.~J., Selsing, J., {et~al.} 2019, Nature, 574, 497–500

\bibitem[{Zhang(2019)}]{Zhang2019}
Zhang, B. 2019, Frontiers of Physics, 14, 64402

\bibitem[{Zhang {et~al.}(2017)Zhang, Xiong, Liao, {et~al.}}]{grb170921C}
Zhang, Y.~F., Xiong, S.~L., Liao, J.~Y., {et~al.} 2017, GRB Coordinates
  Network, 21919

\end{thebibliography}

\end{document}